\documentclass[useAMS]{mn2e}
\usepackage{graphicx}
\usepackage{epsfig}
\usepackage{amssymb}
\usepackage{lscape}
\usepackage{ulem}
\usepackage{txfonts}

\def\kms{km~s$^{-1}$}

\def\c2s{C\,{\sc ii}$^{\star}$}

\def\LIR{L$_{IR}$}

\title[IR luminosities from artificial neural networks] {The infra-red luminosities of $\sim$332,000 SDSS galaxies predicted from artificial neural networks and the Herschel Stripe 82 survey}

\author[Ellison et al.] {Sara L. Ellison$^1$,
  Hossein Teimoorinia$^1$,
  David J. Rosario$^2$,
J. Trevor Mendel$^2$\\
$^1$ Department of Physics \& Astronomy, University
of Victoria, Finnerty Road, Victoria, British Columbia, V8P 1A1,
Canada.\\
$^2$ Max-Planck-Institut fur Extraterrestrische Physik, Giessenbachstrasse, D-85748 Garching, Germany.
}
\begin{document}

\maketitle

\begin{abstract}

The total infra-red (IR) luminosity (\LIR) can be used as a robust
measure of a galaxy's star formation rate (SFR), even in the presence
of an active galactic nucleus (AGN), or when optical emission lines
are weak.  Unfortunately, existing all sky far-IR surveys,
such as the Infra-red Astronomical Satellite (IRAS) and AKARI,
are relatively shallow and are biased towards the highest SFR galaxies
and lowest redshifts.  More sensitive surveys with the Herschel Space
Observatory are limited to much smaller areas.  In order to construct
a large sample of \LIR\ measurements for galaxies in the nearby universe,
we employ artificial neural networks (ANNs), using 1136 galaxies in the 
Herschel Stripe 82 sample as the training set.  The networks are validated
using two independent datasets (IRAS
and AKARI) and demonstrated to predict the \LIR\ with a scatter $\sigma \sim$
0.23 dex, and with no systematic offset.  Importantly, the ANN performs
well for both star-forming galaxies and those with an AGN.
A public catalog is presented with our \LIR\ predictions which can be used
to determine SFRs for 331,926 galaxies in the Sloan Digital Sky Survey (SDSS),
including $\sim$ 129,000 SFRs for AGN-dominated galaxies
for which SDSS SFRs have large uncertainties.  

\end{abstract}


\section{Introduction}

Galaxies evolve through the conversion of gas into stars.  Galactic
mass assembly over cosmic time is therefore largely characterized
by the history of star formation.  In turn, the star formation is modulated by both
internal processes that may both trigger and quench star formation
(e.g. Tan 2000; Ellison et al. 2011; Cheung et al. 2012; Saintonge et al. 2012;
Wang et al. 2012; Mendel et al. 2013; Bluck et al. 2014; Forbes et al. 2014;
Schawinski et al. 2014; Willett et al. 2015)
as well as external gas replenishment from the inter-galactic
medium and mergers (Keres et al. 2005; Sancisi et al. 2008;
L'Huillier, Combes \& Semelin 2012).  Large samples of robust
star formation rates are therefore critical for dissecting the
processes that drive galaxy evolution.

Although galaxy scale star formation rates (SFRs) can be derived at 
many wavelengths, ranging from the X-ray to the radio
(see Kennicutt 1998 and Kennicutt \& Evans 2012 for reviews), the
most common techniques (at low redshifts) use rest-frame ultra-violet (UV) and
optical diagnostics. In the
UV, the unobscured SFR is measured directly from the photons
originating from the young stellar population.  However, UV diagnostics
must usually be combined with data at longer wavelengths in order to
account for dust obscured star formation (e.g. Salim et al. 2007 and references therein).
Optical emission lines suffer less from (but must still be corrected for)
the effects of dust obscuration, and represent one of the most widely used 
methods for measuring the SFR in the local universe.
Such techniques either measure the strength of Balmer recombination lines (e.g.
Kennicutt et al. 1994),
which are a direct consequence of ionization by UV photons, or make use of
empirical calibrations with forbidden lines such as [O~II] (e.g. Kennicutt 1998). 
The main limitation of optical emission lines is contamination
from ionizing processes that are not linked to young stars, such as interstellar shocks,
active galactic nuclei (AGN) or planetary nebulae.  The calibration between
[O~II] luminosity and SFR is also dependent on dust extinction and metallicity;
if these parameters are not accounted for, additional scatter arises in the SFR calibration 
(Kewley, Geller \& Jansen 2004).  The measurement of
the far infra-red (FIR) continuum, which probes emission from warm dust heated
by star formation, provides a third complementary SFR measure.  Beyond the
mid-IR, contributions to the FIR from AGN should be negligible and can
yield a robust measure of the total SFR  (e.g. Netzer
et al. 2007; Buat et al. 2010; Hatziminaoglou et al. 2010;
Mullaney et al. 2011).

Despite its advantages, the application of FIR star formation rate
diagnostics in the nearby universe has remained fairly limited, and the
largest SFR samples currently available tend to rely on optical emission lines
and the UV (Brinchmann et al. 2004; Salim et al. 2007; Gunawardhana et al.
2011).
This limitation has been due largely to the typically low sensitivity
and low spatial resolution of FIR observing facilities.  The former
issue limits detections to only the highest SFR galaxies and the latter
issue means that measurements may be contaminated by either other galaxies
in the telescope beam or by Galactic cirrus (e.g. Takeuchi et al. 2010). The Herschel Space
Observatory (Pilbratt et al. 2010; hereafter, simply Herschel) represents a dramatic
improvement in both spatial resolution and in sensitivity.  There are several
extra-galactic surveys that have been conducted with Herschel, whose operations ceased
in 2013.  The largest is the
Herschel Astrophysical Terahertz Large Area Survey (H-ATLAS, Eales et al.
2010), which covered over 550 sq. degrees and is predicted to yield
many tens of thousands of detections of galaxies at $z<0.3$.  Indeed,
in the Science Demonstration Phase (SDP), H-ATLAS covered 16 sq. degrees
and detected $\sim$ 3300 galaxies at $z<0.2$ (Dariush et al. 2011).
A review of FIR galaxy surveys, with an emphasis on Herschel's impact,
is given in Lutz (2014).

Whilst Herschel can provide robust SFRs for galaxies that are actively
forming stars, it reaches its sensitivity limit for a SFR of $\sim$ 1
M$_{\odot}$ yr$^{-1}$ at z=0.08 (Rosario et al. 2015).  For example,
the H-ATLAS survey finds that the 250 $\mu$m
detection rate of low redshift galaxies with $r$-band magnitudes
brighter than 18 is $\sim$30 per cent (Dariush et al. 2011).  Similar
detection fractions are found from other Herschel galaxy surveys,
such as the Herschel Multi-Tiered Extragalactic Survey (HerMES,
Buat et al. 2010; Hatziminaoglou et al. 2010) 
and the Herschel Stripe 82 survey (Viero et al. 2014; Rosario et al. 2015).
The importance of Herschel's detection 
limit is poignantly demonstrated when a comparison is made with optical SFRs.
In general, the SFRs of \textit{actively star forming} galaxies derived from 
optical emission lines are in excellent agreement with those determined 
from Herschel FIR photometry (Dominguez Sanchez et al. 2012; Rosario
et al. 2015).  However,
for galaxies with an AGN contribution, or weak emission lines there is an
apparent offset between the optical SFRs in the Sloan Digital Sky Survey
(SDSS) and those determined from the FIR (Fan et al. 2013; Matsuoka
\& Woo 2015).  For these
galaxies, the SDSS SFRs are derived from the 4000 \AA\ break, which
correlates well with specific SFR in star-forming galaxies (see Brinchmann
et al. 2004 for a complete description of this method), but uncertainties
in the calibration lead to large uncertainties on the resulting SFR.
Rosario et al. (2015) have shown that the apparent under-estimate of
SDSS SFRs, relative to that derived from \LIR, for weak emission line galaxies and AGN
is due to a combination of the limiting SFR detectable by Herschel,
and the relatively large uncertainties on the optical SFRs for
weak line and AGN galaxies. Therefore, despite its detection
limit, Rosario et al. (2015) have demonstrated that Herschel
is able to provide robust SFRs for classes of galaxies (such as those
with low SFRs, or AGN contributions) that frequently have large uncertainties
in their optical SFRs.  

The promise of FIR star formation rates is currently limited
by the size of the Herschel samples available.  The final 
release of H-ATLAS data will be a dramatic step forward in this regard,
but its sky coverage is still less than 10 per cent of the SDSS Data
Release 7 (DR7)
Legacy area.  There are no imminent prospects for expanding deep
FIR surveys; the next major IR facility will be the joint 
Japanese/European SPICA
observatory, not due for launch for another decade.  In this paper
we investigate  the possibility of using existing Herschel data
as a training set for machine learning techniques that could be
used to predict the \LIR\ for a large sample of nearby galaxies. 
We have previously used artificial neural networks (ANN) to predict
emission line fluxes for galaxies in the SDSS (Teimoorinia \& Ellison
2014) and for ranking the parameters that drive galaxy quenching
(Teimoorinia, Bluck \& Ellison 2015).

The current paper is organized as follows.
In Section \ref{sec-data} we describe the various IR observational
samples and catalogs used in this work, for either training or verification
of the ANN, the determination of \LIR, and consistency checks between
these catalogs.  In Section \ref{sec-ann} we outline the basic
ANN training procedure; the validation of the network is described
in Section \ref{sec-valid}.  The trained network is applied to the
SDSS in Section \ref{sec-sdss}, resulting in a catalog of $\sim$
332,000 galaxies with predicted IR luminosities, which is made publically
available with this paper.

\section{Infra-red observational samples and catalogs}\label{sec-data}

In this work we will make use of data from three separate
spacecraft that collected data in the far-IR: the
Infra-red Astronomical Satellite (IRAS, Neugebauer et al. 1984),
AKARI (Murakami et al. 2007) and
Herschel (Pilbratt et al. 2010).  The first two of these missions were designed
to perform all sky surveys that have public photometry
available in their archives. However, as we will demonstrate later
in this Section, AKARI and IRAS are limited to galaxies with
relatively high SFRs.  We therefore use data from the more sensitive
Herschel mission as our training set.  In contrast to AKARI and IRAS,
Herschel operations combined targeted observations with survey science, 
including the H-ATLAS key program (Eales et al. 2010), of which
only the 16 sq. degrees of the SDP are currently publically available.
The public catalogs for the HerMES fields are the next largest dataset,
covering a total of $\sim$ 100 sq. deg.  However, the different
fields are observed to different depths (e.g. Fig 1 of Lutz 2014; see
Oliver et al 2012 for a complete description of the multi-tiered approach),
yielding a dataset with inhomogeneous detection thresholds, which is 
sub-optimal for training an ANN.
Therefore, in order to train the network with the largest, homogeneous
public dataset, which is critical for a robust learning pattern, we use
the results from the Herschel Stripe 82 survey (Viero et al. 2014)
which covers 79 sq. degrees.  

Previous works have computed the total
IR luminosity from these public catalogs. We review several
of these works in the sub-sections below, but refer the interested
reader to the original survey papers for full details on
photometry and spectral fitting.  However, since the ANN
procedure involves training on one dataset, followed by validation
with independent datasets\footnote{When one large 
  training set is available, it is common practice to train on
  a fraction of the data, and use the remainder for the independent
  validation.  However, our training set is rather limited in number
  (N=1136), so that we risk compromising the training procedure if we
  reduce its input.}, homogeneity between the catalogs is
paramount. Systematic differences between \LIR\ derivations,
e.g. due to different far-IR templates, will compromise the
performance of the ANN predictions.  Therefore, although we
make use of the archival catalogs of IR photometry, we will
derive all of our own IR luminosities.

We begin with a brief review of the characteristics of the IR data, the
public catalogs and matches to the SDSS. We then describe the techniques
applied homogeneously to the photometric catalogs to determine the total
infra-red luminosities.

 \begin{figure}
\centering
\includegraphics[width=7.cm,height=5.5cm,angle=0]{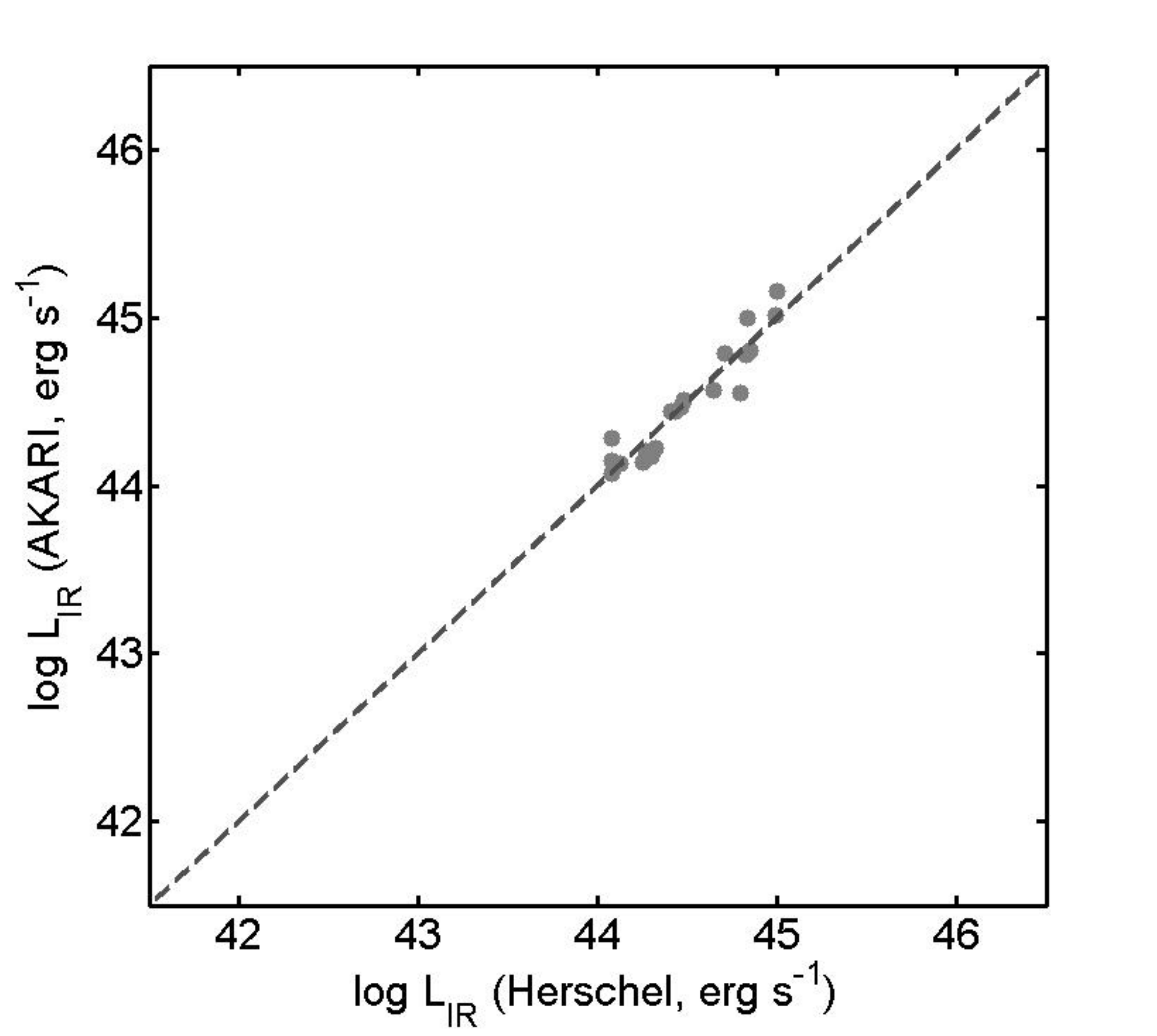}
\includegraphics[width=7.cm,height=5.5cm,angle=0]{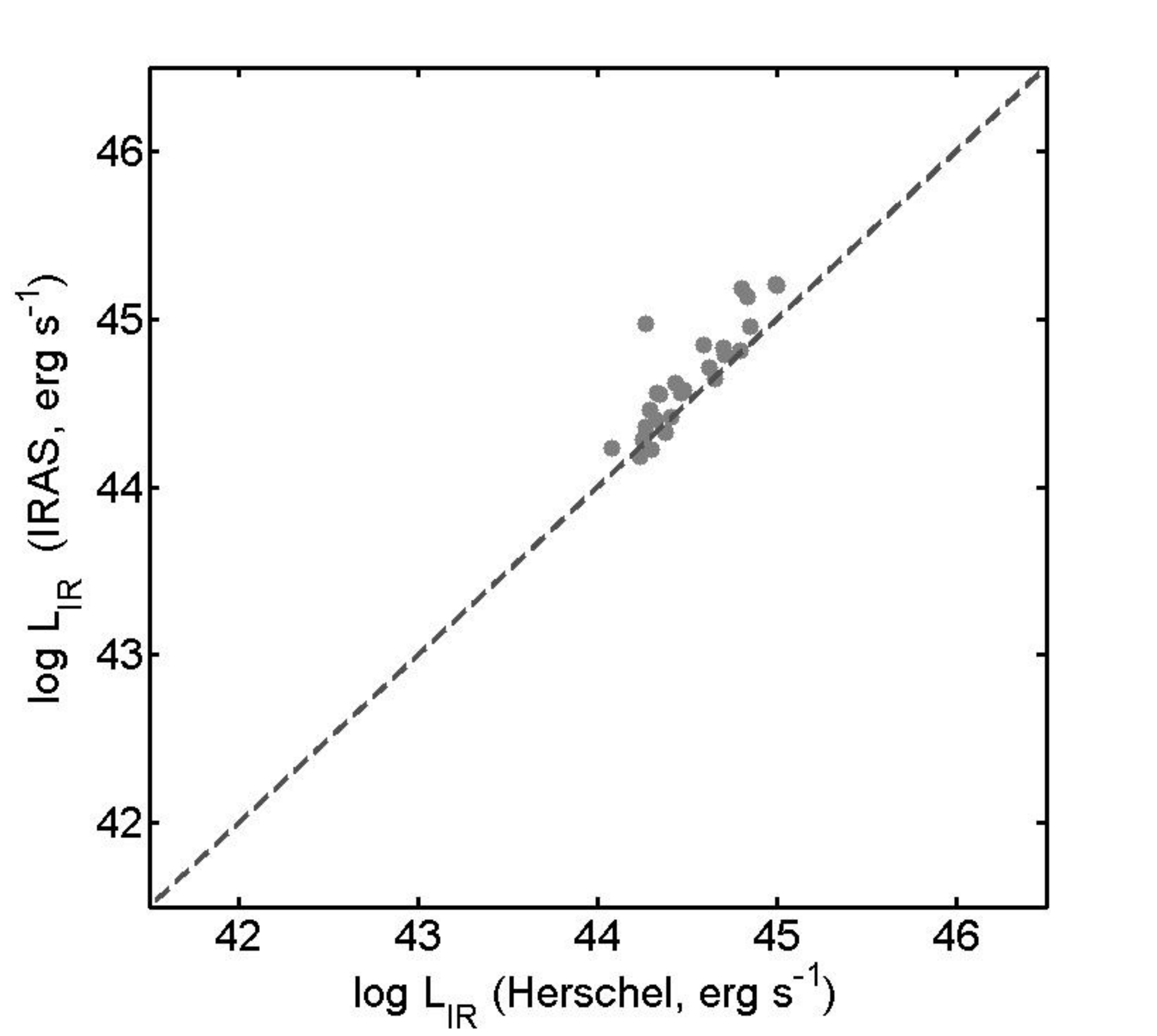}
\includegraphics[width=7.cm,height=5.5cm,angle=0]{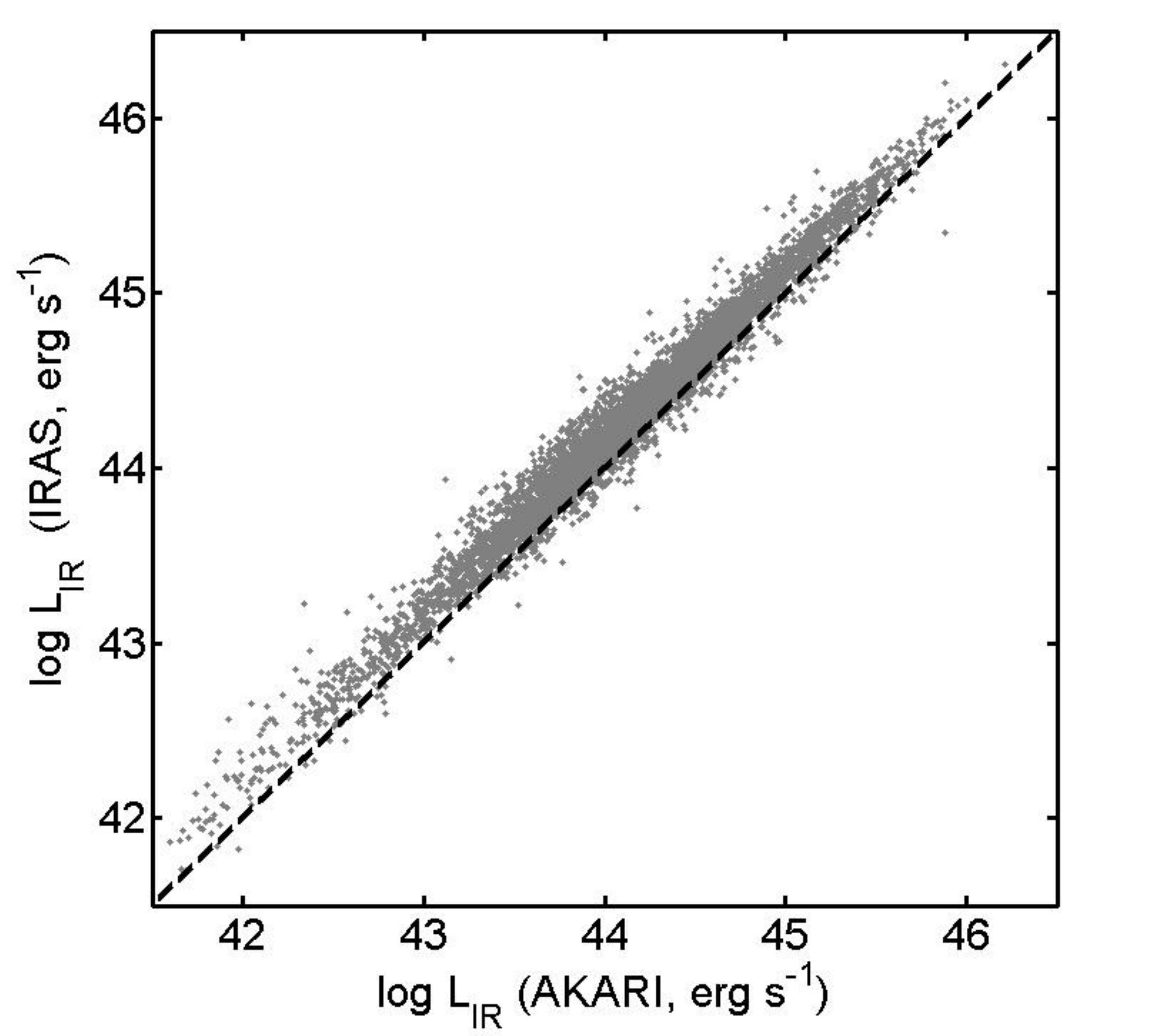}
\caption{Comparison between derived \LIR\ for galaxies in common between the
  three catalogs considered in this work.  Whereas AKARI and Herschel are
  in good agreement, the infra-red luminosities from IRAS are offset from the other two catalogs
by $\sim$ 0.15 dex}
\label{fig-compare-cats}
\end{figure}

 \begin{figure}
\centering
\includegraphics[width=7.cm,height=5.5cm,angle=0]{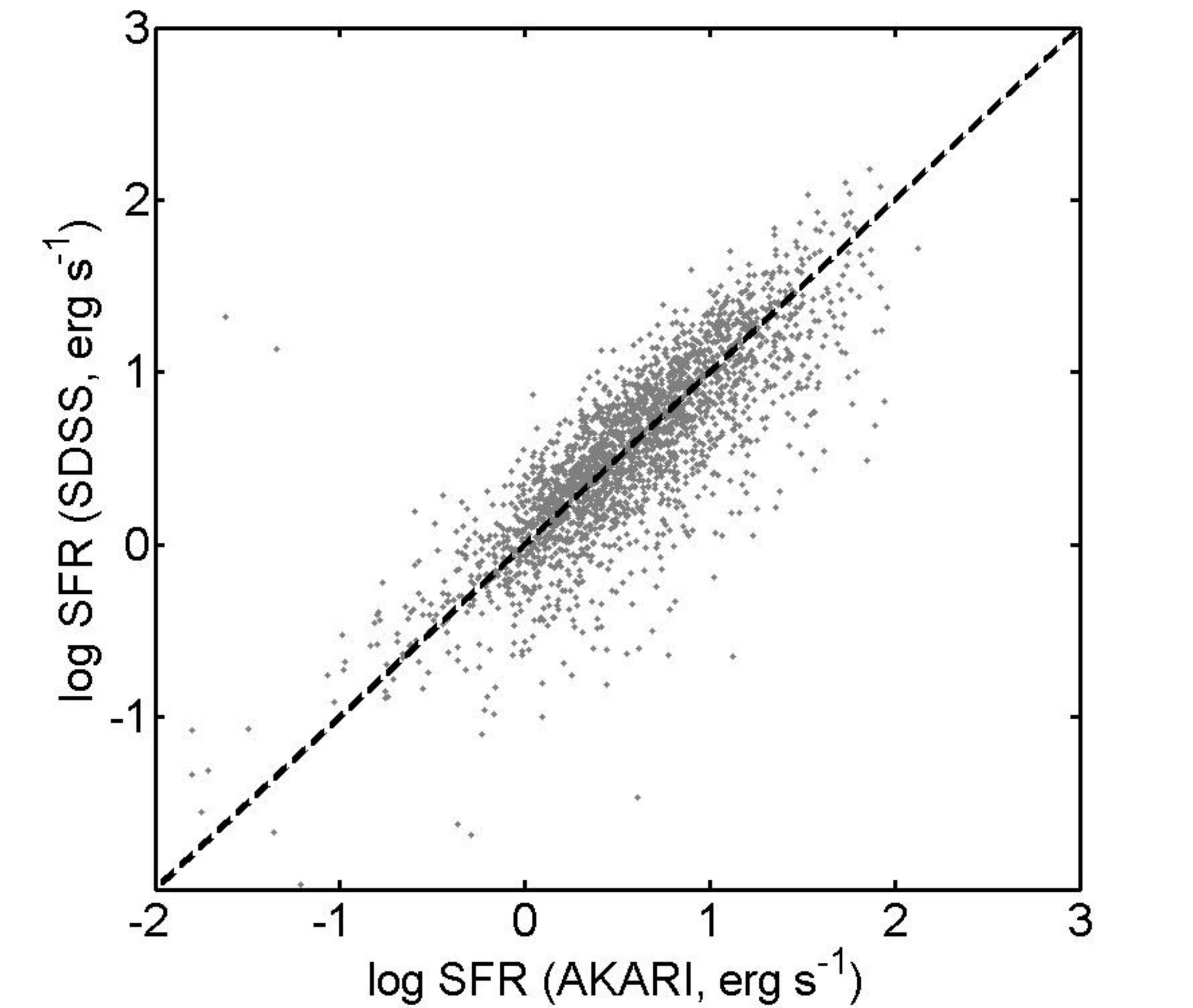}
\includegraphics[width=7.cm,height=5.5cm,angle=0]{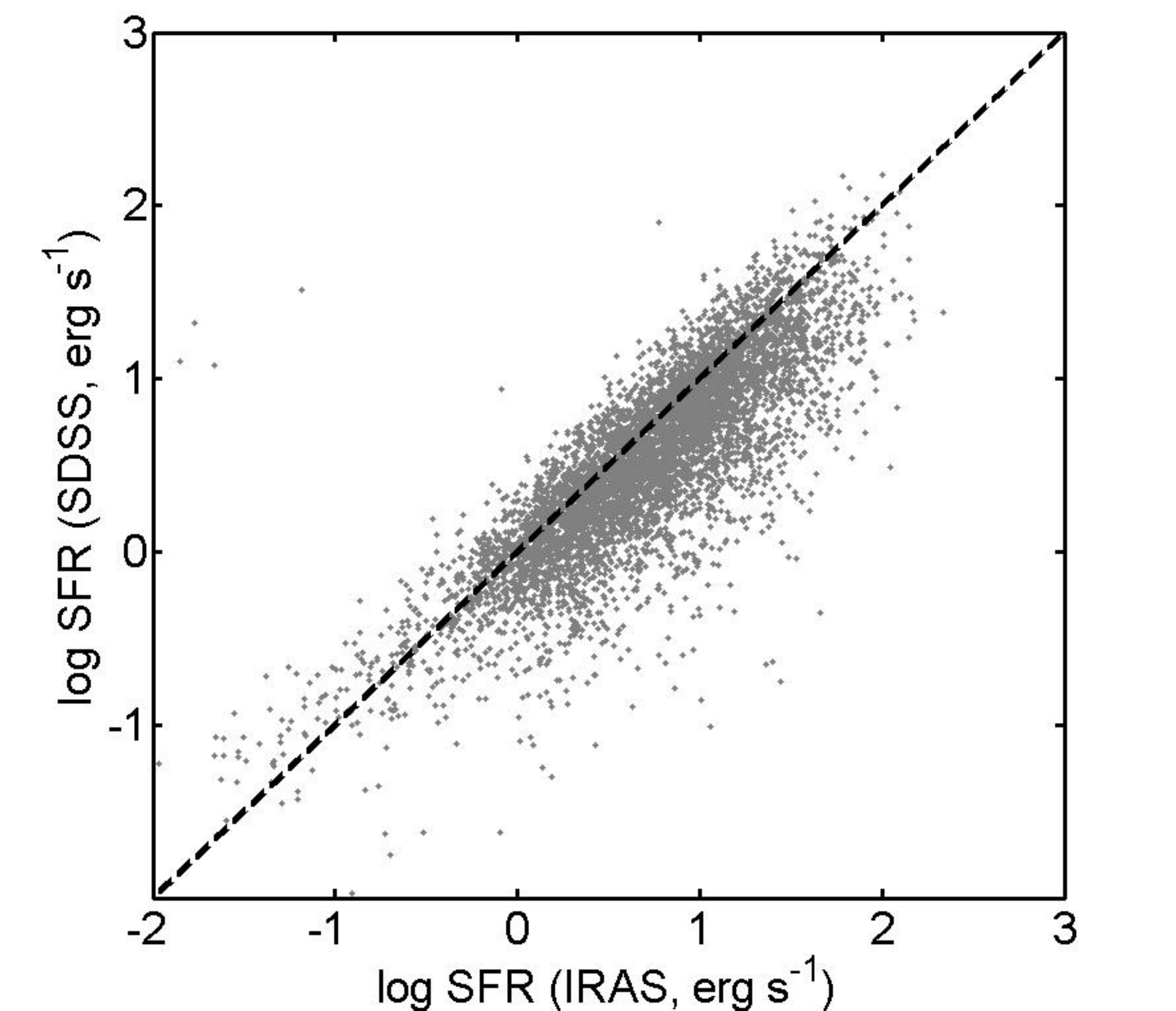}
\includegraphics[width=7.cm,height=5.5cm,angle=0]{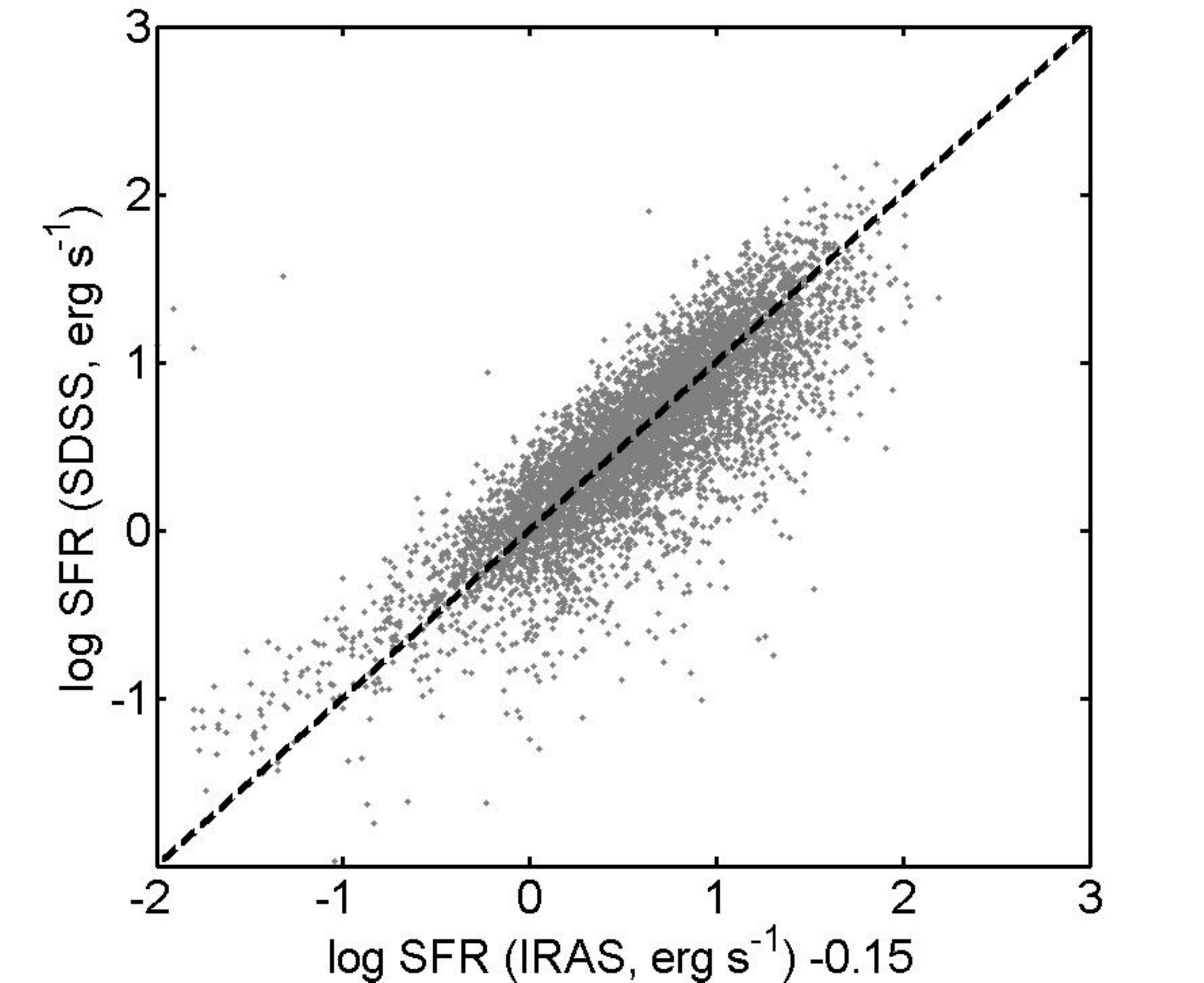}
\caption{Comparison between the aperture corrected total
SFRs (star-forming galaxies only) measured
  from the SDSS spectra and values inferred from \LIR\ (Eqn. \ref{eqn-sfr})
  for AKARI (top panel) and IRAS (middle panel).  The median offset of the IRAS
  fluxes derived from both the SFRs (this Figure) and \LIR\ (Figure
  \ref{fig-compare-cats}) is 0.15 dex.  The lower panel shows the SFRs derived
from IRAS after this correction has been applied.  A comparison between SFRs inferred for Herschel detected galaxies and the SDSS values is presented in Rosario et al. (2015), where an excellent agreement is found for star-forming galaxies.}
\label{fig-compare-cats-sfr}
\end{figure}

\subsection{The IRAS All Sky Survey}

IRAS (Neugebauer et al. 1984) surveyed the sky in four bands at
12, 25, 60 and 100 $\mu$m, with a spatial resolution ranging from approximately
0.5 arcmin to 2 arcmin at 12  and 100 $\mu$m respectively.  Numerous works have matched
galaxies in the SDSS to IRAS catalogs in order to study the IR properties
of low redshift galaxies (e.g. Goto 2005; Wang et al. 2006; Obric et al.
2006; Wang \& Rowan-Robinson 2009; Hwang et al. 2010, 2011; Wang et al. 2014).
In this work, we use the catalog of Hwang et al. (2010), who matched
the IRAS Faint Source Catalogue
(FSC, Moshir et al. 1992), which contains 173, 044 IR sources at
$| b | >$ 10 deg, with the SDSS DR7.
As described by Hwang et al. (2010), the positional uncertainties in
the IRAS data are heterogeneous and depend on the scan direction.
The positional uncertainty can be as small as 1 arcsec, ranging up
to almost 1 arcmin.  In order to match the IRAS detected sources
with the SDSS, Hwang et al. (2010) used the individual source error
ellipses and identified successful matches within 3 $\sigma$ of
this positional uncertainty.  This procedure yielded multiple matches
for 10 per cent of IRAS detections, in which case the closest optical
match was adopted.  There are a total of 14,334 galaxies matched between
SDSS and IRAS with 60 $\mu$m detections.  Since the determination of
\LIR\ (Section \ref{lir_fit}) includes mid-IR\footnote{Unless otherwise 
stated, all mid-IR photometry in this paper comes from WISE.  } 
photometry from  the Wide Field Infra-red Survey 
Explorer (WISE, Wright et al. 2010), we limit our sample to $z<0.3$
in order to avoid possible contamination of polyaromatic hydrocarbons (PAHs).
After this limiting redshift is imposed, our final IRAS sample consists 
14,255 galaxies for which \LIR\ is computed.

\subsection{The AKARI/FIS All Sky Survey}\label{akari_sec}

The Far Infra-red Surveyor (FIS, Kawada et al. 2007) on board the AKARI satellite
(Murakami et al. 2007) conducted an all sky survey at 65, 90, 140 and
160 $\mu$m wavelengths.  The nominal point spread functions of FIS are
37, 39, 58 and 61 arcseconds in these four wavebands, respectively (Kawada
et al. 2007).  The FIS Bright Source Catalog 90 $\mu$m detected sources
were matched to the SDSS DR7 by Hwang et al. (2011). A matching tolerance
of 12 arcsec was adopted, which corresponds to AKARI's 2 $\sigma$ 
positional uncertainty.  If multiple galaxies are matched, the closest
one is adopted.

We once again make use of the positional matches of Hwang et al.
(2011) but compute our own infra-red luminosities.  Nonetheless, we have
compared our derived \LIR\ values with those of Hwang et al. (2011), who
used the Chary \& Elbaz (2001) templates.
In general there is very good agreement
in the two sets of infra-red luminosities (the same is also true of
a comparison between our IRAS fluxes and those of Hwang et al. 2010),
where our \LIR\ determinations using the Dale \& Helou (2002)
templates are higher
than those of Hwang et al. by $\sim$ 0.03 dex on average.  This is
consistent with previous works (e.g. Goto et al. 2011) who have also
concluded that robust \LIR\ determinations can be obtained independent
of the choice of template.  In total, there are 6965 galaxies
with 90 $\mu$m detections in the matched SDSS catalog; after the
$z<0.3$ cut, this results in a final AKARI sample of 6957
galaxies for which \LIR\ is derived.

\subsection{The Herschel/SPIRE Stripe 82 Survey}

The Herschel Stripe 82 survey (Viero et al. 2014) covers a total
area of 79 deg$^2$ along an equatorial strip ($-2 < \delta < +2$)
between 13 and 37 degrees in right ascension.  The survey was conducted
by the SPIRE instrument (Griffin et al. 2010)
at 250, 350 and 500 $\mu$m, with a
spatial resolution of 18.2, 25.2 and 36.3 arcseconds respectively.
The catalog is 250 $\mu$m selected, to a depth of $\sim$ 30 mJy
(3$\sigma$), which corresponds to a SFR of 1 M$_{\odot}$/year at $z=0.08$. 

Rosario et al. (2015) matched the Herschel Stripe 82 catalog with galaxies
present in the SDSS DR7 Main Galaxy Sample whose
extinction corrected Petrosian $r$-band magnitudes are brighter than 17.77 and
whose redshifts are in the range 0.04 $< z<$ 0.15. 3319 SDSS galaxies
satisfying these criteria lie within the Herschel Stripe 82 footprint.
The two catalogs were matched with a positional tolerance of 5 arcseconds,
yielding 1349 matches; an IR detection rate of 40\%.  Rosario et al. (2015)
also matched the SPIRE/SDSS detections with the public all sky survey conducted
by WISE, with
the requirement of a S/N$>$ 5 in at least one of the 4 bands (3.4, 4.6, 12 and 22
$\mu$m) and positional tolerance of 2 arcsecs. 

\subsection{Derivation of \LIR}\label{lir_fit}

The mid- and far-IR photometry of galaxies from WISE and each of the FIR surveys 
described above, respectively, were fit with the galaxy SED templates of Dale and Helou (2002). This process is described in detail by Rosario et al. (2015),
but we review the salient details here.  IR luminosities 
were derived by integrating the best fit templates over rest-frame 8-1000 $\mu$m. 
The minimum requirement for a fit was a S/N $>$ 3 detection in the selection band of the survey: IRAS 60 $\mu$m, 
AKARI 90 $\mu$m, or Herschel/SPIRE 250 $\mu$m. If a galaxy had S/N$>$2 photometric measurements in other bands from 
the same survey, these were also used to constrain the fits. The exception is the IRAS 12 $\mu$m band which we do not use, due to the possibility of contamination by AGN.  In the case of Herschel galaxies, WISE 22 $\mu$m 
photometry was used where available to constrain the short-wavelength part of the IR SED. We required that the 
additional bands covered rest-frame wavelengths redwards of 20 $\mu$m to avoid any strong effects from PAHs 
or AGN-heated dust.

Since the selection band is also the most sensitive in each of these surveys, a substantial fraction of 
galaxies were only detected in this single band. In these cases, we estimated \LIR\ by scaling the photometry 
to a fixed IR template, corresponding to an M82-like SED (see Rosario et al. 2015 for more details about rationale 
for this choice of template.)

\subsection{Calibration consistency}

We now have in hand three separate IR catalogs that have been matched
to the SDSS and for which we have computed the \LIR. 
Before discussing how the ANN training and validation sets are selected,
we must first confirm that the different IR datasets are consistently
calibrated.  Despite the adoption of a uniform spectral template fitting
procedure, the measurement of the input photometry is heterogeneous, the
compilation of catalogs is performed at different wavelengths and the
range in spatial resolution between instruments may also result in varying
degrees of contamination or SDSS matching errors.  

We perform the consistency checks between IR catalogs in two ways.  First, we identify
the galaxies that are in common between the three surveys, resulting in 20 galaxies matched
between Herschel Stripe 82 and AKARI, 26 galaxies between Herschel and IRAS and 5860
between IRAS and AKARI. In Figure \ref{fig-compare-cats} we compare the \LIR\
of these common galaxies.    For the galaxies common between the AKARI and
Herschel Stripe 82 catalogs, the \LIR\ are in excellent agreement, with no systematic
offset (Fig.  \ref{fig-compare-cats}, top panel).  However, the common galaxies in
Herschel and IRAS are slightly offset from the 1:1 relation by a mean value of 0.14 dex  
(Fig. \ref{fig-compare-cats}, middle panel), although the number
of galaxies is small.  The offset is more obvious between AKARI and IRAS, thanks
to the larger dataset
(Fig.  \ref{fig-compare-cats}, bottom panel), in which the mean offset for $\sim$ 6000
galaxies is 0.15 dex.

Due to the relatively small overlap between the Herschel Stripe 82 SDSS sample and either
AKARI or IRAS, we perform a second test of the consistency between catalogs.
In Rosario et al. (2015) we showed that, for star forming galaxies, the SFRs inferred
from the Herschel \LIR\ are in excellent agreement with the values derived from the aperture
corrected SDSS spectra.  We therefore similarly use the SDSS total SFRs to verify the \LIR\ values
of AKARI and IRAS.  We adopt the conversion between \LIR\ and SFR from Kennicutt (1998),
with an adjustment for a Chabrier initial mass function (IMF):

\begin{equation}\label{eqn-sfr}
\log  L_{IR} = \log SFR + 43.591
\end{equation}

SDSS SFRs are taken from the public MPA/JHU catalogs
\footnote{http://home.strw.leidenuniv.nl/\~{}jarle/SDSS/}.  The techniques
used to determine total SFRs for SDSS galaxies
are described in detail by Brinchmann et al. (2004)
and Salim et al. (2007), but we review the salient details here.  The
fibre spectra are fit with models from Charlot \& Longhetti (2001), which
combine stellar continuum templates with emission lines simulated by
the Cloudy software (Ferland et al. 1998).  The models are run over large grids
of input parameters, including dust attenuation, metallicity and star formation
rate.  Likelihood distributions are constructed using a Bayesian approach,
from which the median is taken as the `best' value.  Errors on SFRs are derived
from the full probability distribution functions.  Although the models
fit the entire SDSS spectrum, the derived SFRs are dominated by the strength of
the Balmer emission lines.    However, Brinchmann et al. (2004) note
that contributions to emission lines from processes not related to star formation
will lead to erroneous SFRs.  Such processes include shocks, planetary nebulae
and AGN.  Brinchmann et al. (2004) therefore identify galaxies with either
a minor (`composite') or dominant AGN contribution using emission line fluxes
and the diagnostic diagrams of Kewley et al. (2001) and Kauffmann et al. (2003).
For both the composite and AGN galaxies, the SFR is determined from the
correlation between D$_{4000}$ and specific SFR (SSFR), as measured in the
star-forming population.  Errors on SFRs derived from D$_{4000}$ are
computed by convolving the likelihood distribution of a given galaxy's
D$_{4000}$ with that of the relation between SSFR and D$_{4000}$ in
star forming galaxies.  For all galaxy classes, the `fibre' SFRs are converted 
to total SFRs by assessing the photometry outside of the aperture.

In Figure \ref{fig-compare-cats-sfr} we plot the SFRs derived (using Eqn. \ref{eqn-sfr})
from the observed (AKARI or IRAS) infra-red luminosities compared with the aperture 
corrected values derived from the SDSS spectra for star forming galaxies (Kauffmann et al. 
2003).  In the top panel, we see that the bulk of the SDSS star-forming galaxies
in the AKARI catalog follow the 1:1 relation.  There is a minority of galaxies that
seem to have large AKARI SFRs compared to the SDSS values; these may be cases where the
relatively large AKARI beam is picking up additional IR flux from a neighbouring
object.  We check this hypothesis by visually inspecting the SDSS images for all
galaxies with log SFR(AKARI) $-$ log SFR(SDSS) $>$ 0.6 dex and find that the majority are
indeed in close pairs, or have multiple companions.  Additionally, we rule out
higher extinctions in these galaxies as the explanation for the SFR offsets,
since these galaxies have unremarkable values of E(B$-$V), as derived from
the Balmer decrement.
In the middle panel of  Figure \ref{fig-compare-cats-sfr}, the same minority 
population with large offsets is seen for the IRAS detections.  However, even the bulk
of the IRAS SFRs seem to be slightly
over-estimated relative to the SDSS values, consistent with the results of
Figure \ref{fig-compare-cats}.  Once again, the median offset required to bring
the IRAS infra-red luminosities into agreement with the SDSS SFRs is 0.15 dex.
In the lower panel of Fig.
\ref{fig-compare-cats-sfr} we show that, after a downwards
correction of 0.15 dex applied uniformly to the IRAS infra-red luminosities,
the IRAS \LIR\ values are now, on average,
in good agreement with the SFRs of star-forming galaxies in the SDSS.

The presence of an offset between AKARI and IRAS \LIR\ has been previously
reported by Takeuchi et al. (2010), who constructed full SEDs extending
from the UV to the far-IR using GALEX, SDSS, 2MASS, IRAS and AKARI.
Takeuchi et al. (2010) demonstrate that the IRAS 60 and 100 $\mu$m
fluxes consistently over-estimate the AKARI 65 and 90 $\mu$m photometry,
suggesting that this is due to increased contamination, e.g. by Galactic
cirrus or nearby sources, in the large IRAS beam (see also Takeuchi
\& Ishii 2004 and Jeong et al. 2007).

In summary, we find that there is good agreement between the \LIR\
determined from Herschel, AKARI and that inferred from the SDSS SFRs, but that IRAS IR
luminosities are systematically over-estimated.  When IRAS \LIR\ s are used in
the remainder of this paper, they are always corrected downwards by 0.15 dex.

 \begin{figure}
\centering
\includegraphics[width=8cm,angle=0]{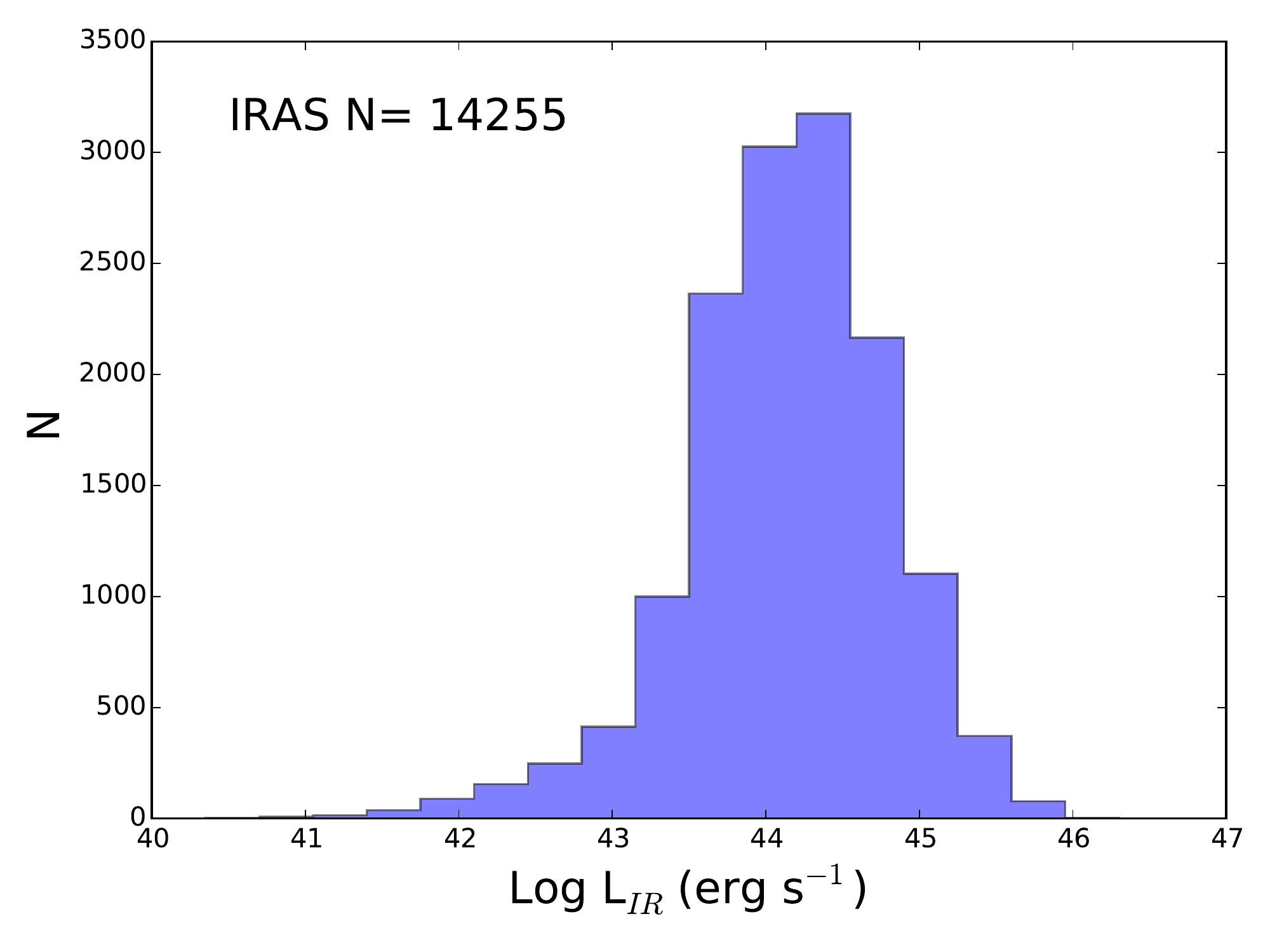}
\includegraphics[width=8cm,angle=0]{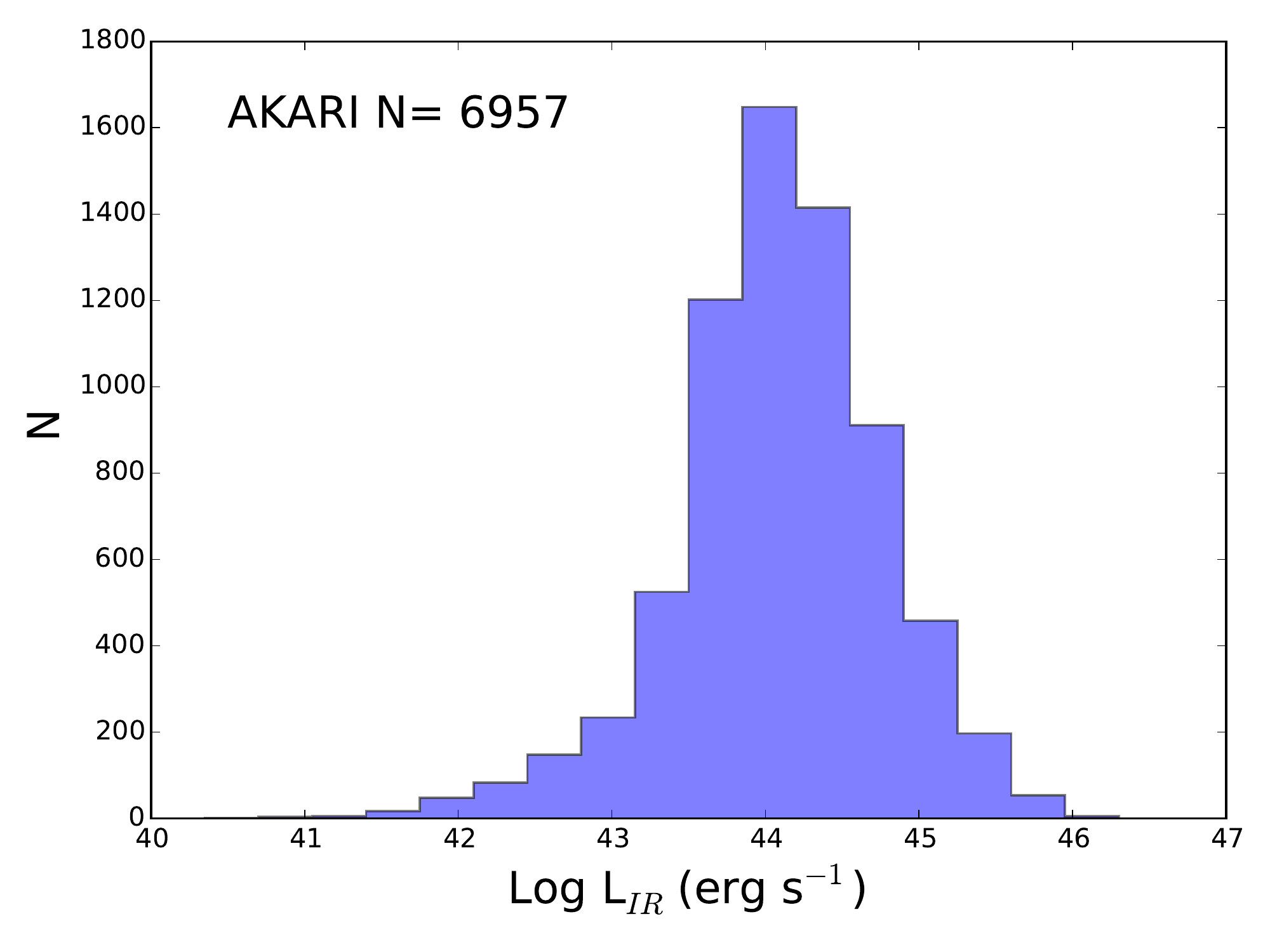}
\includegraphics[width=8cm,angle=0]{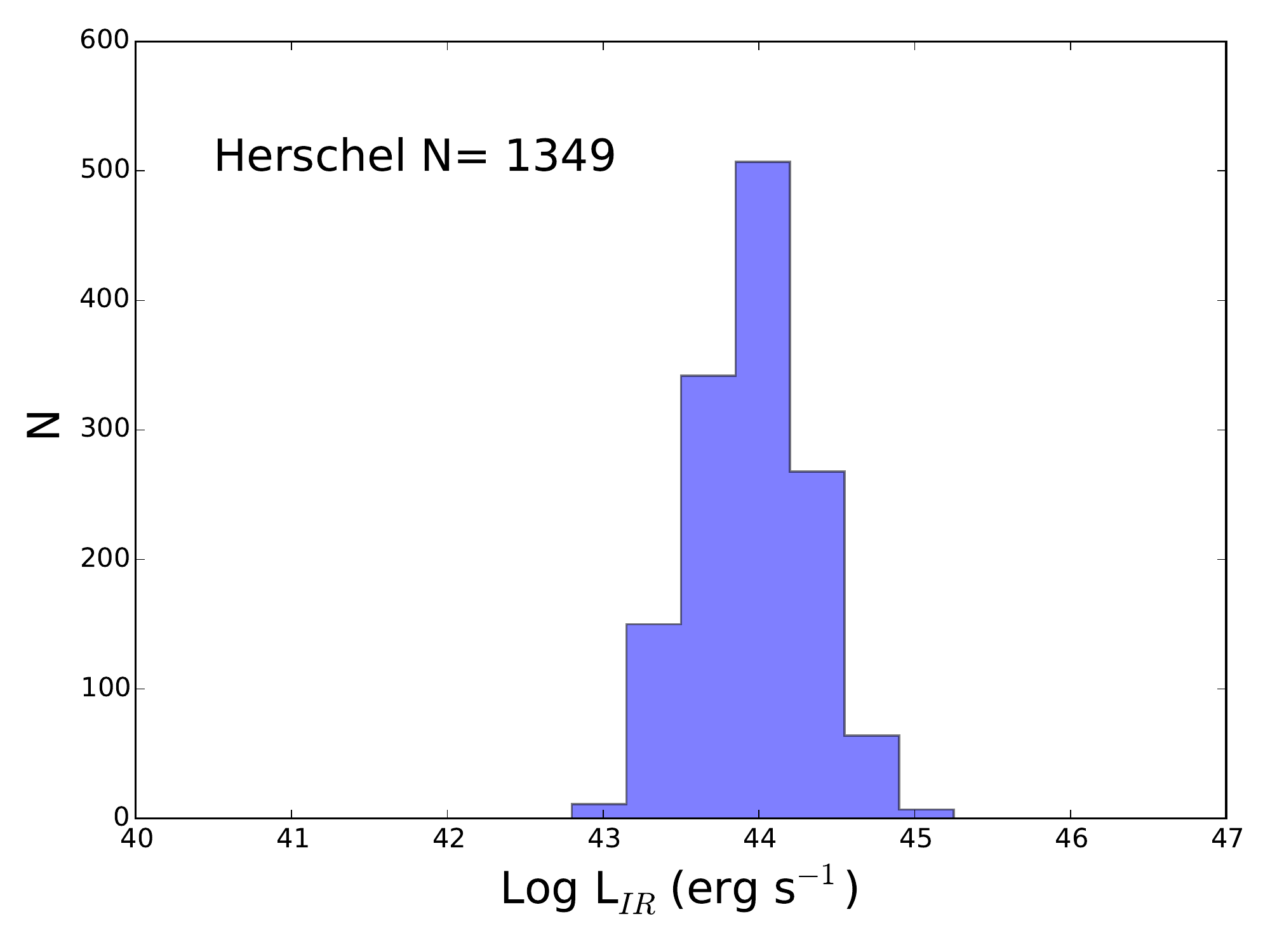}
\caption{The distribution of the measured \LIR\ values for the SDSS matched galaxies
  in the IRAS (top panel), AKARI (middle panel) and Herschel Stripe 82 (lower panel)
datasets.}
\label{fig-lir_hist}
\end{figure}

 \begin{figure}
\centering
\includegraphics[width=8cm,angle=0]{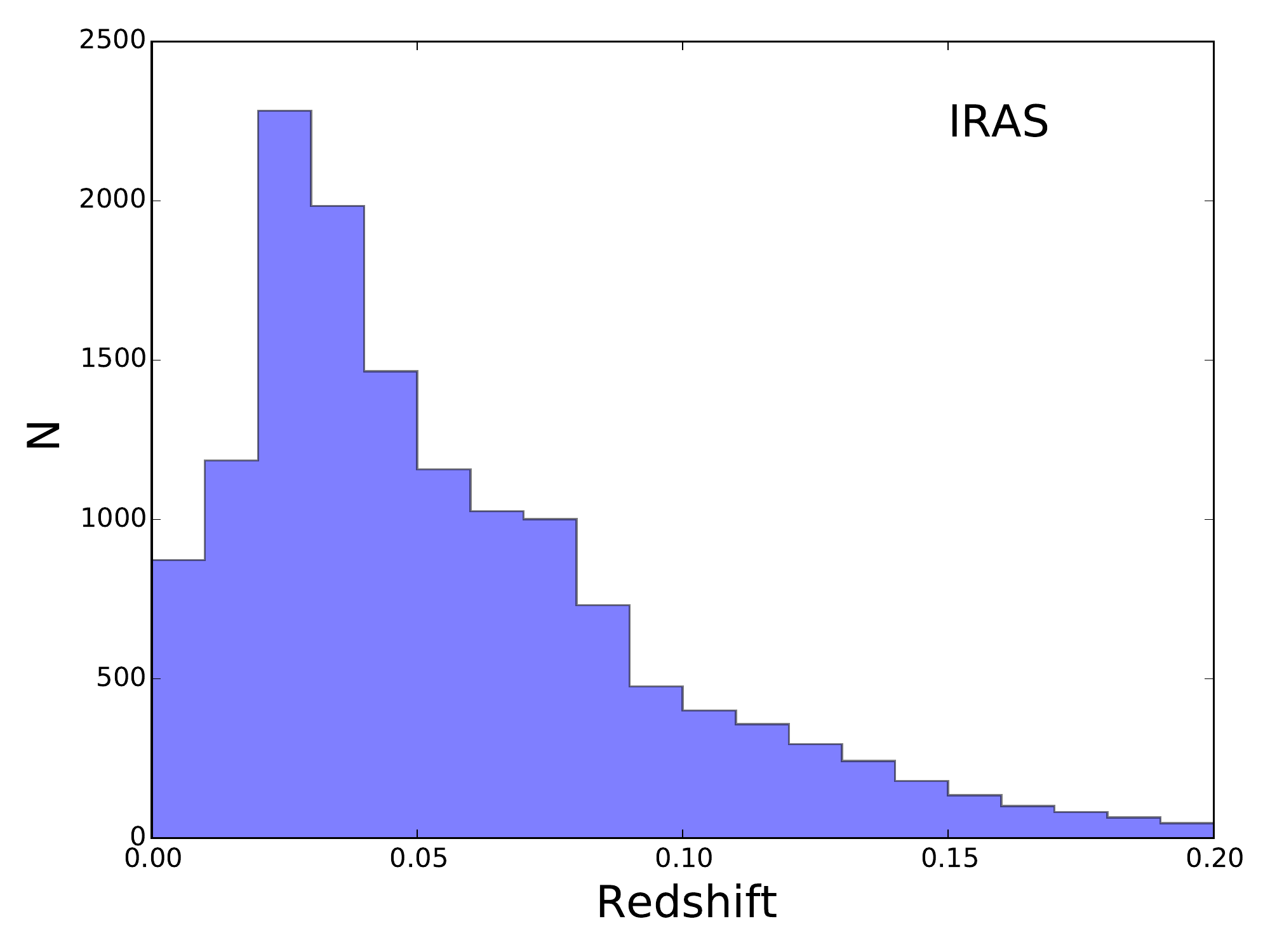}
\includegraphics[width=8cm,angle=0]{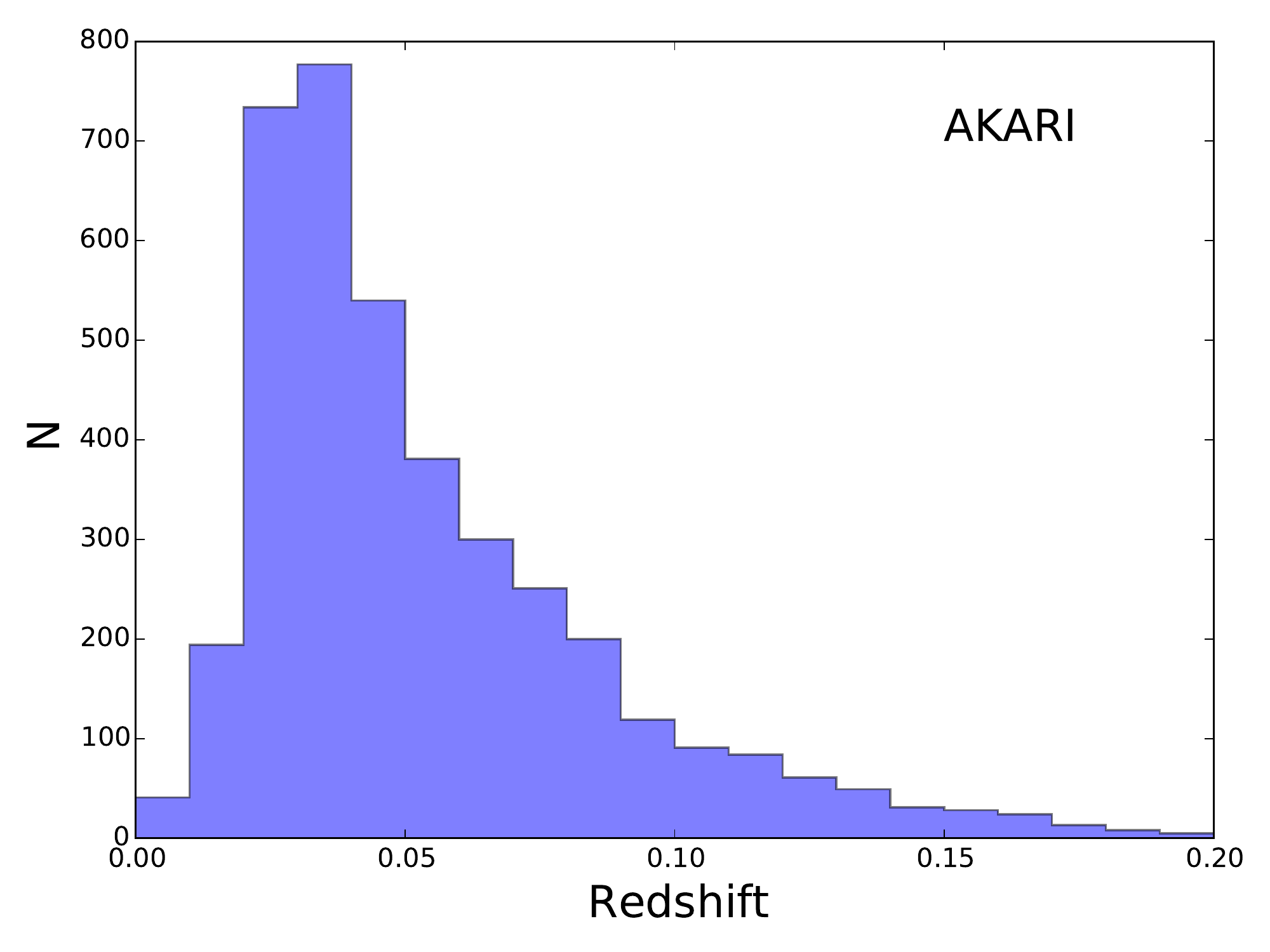}
\includegraphics[width=8cm,angle=0]{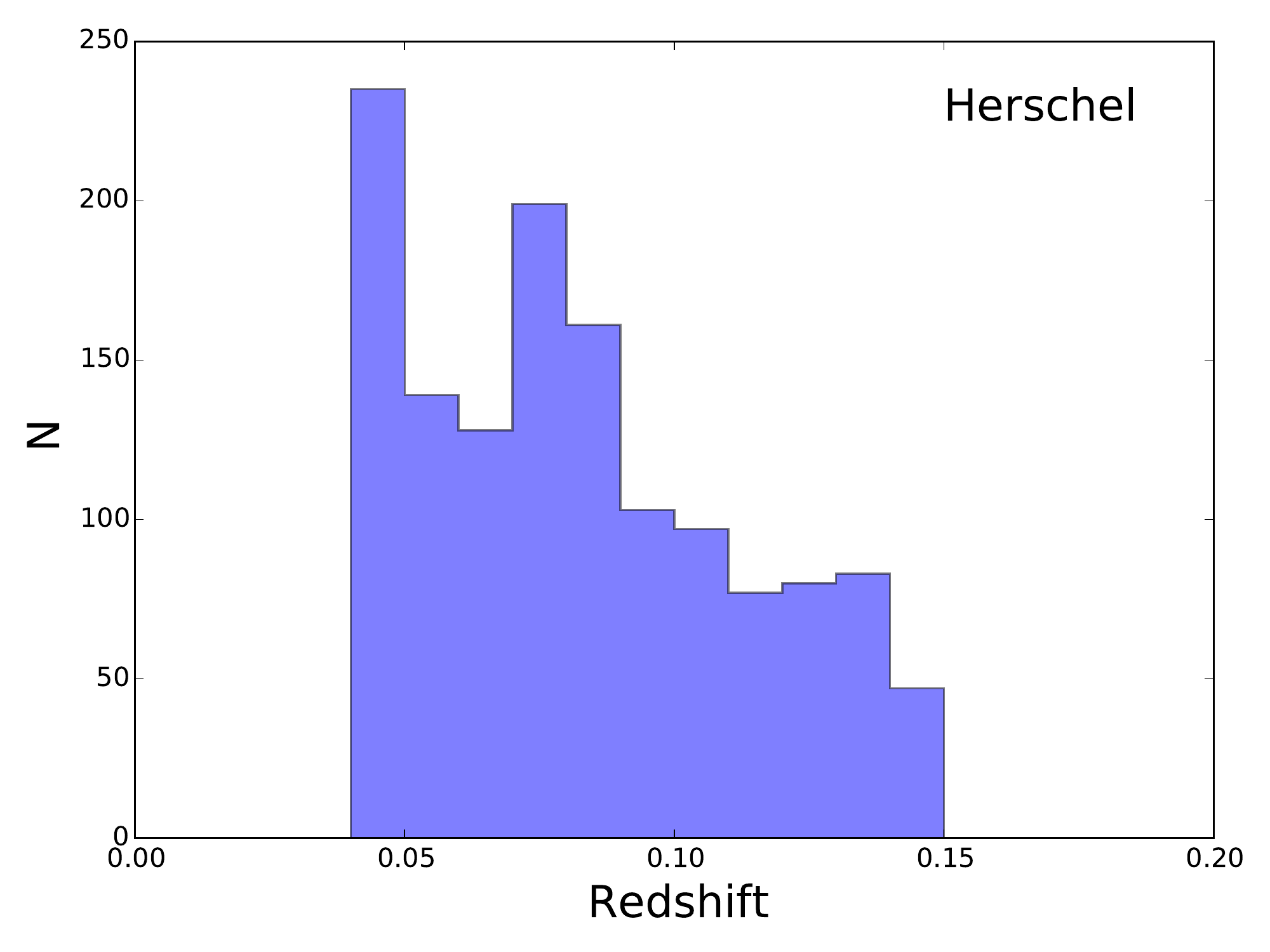}
\caption{The distribution of redshifts for the SDSS matched galaxies
  in the IRAS (top panel), AKARI (middle panel) and Herschel Stripe 82 (lower panel)
datasets.}
\label{fig-zed_hist}
\end{figure}

\subsection{Survey comparisons}\label{sec_compare}

Having established consistency between the \LIR\ calibrations of the
three surveys used in this work, it is now useful to compare their relative
properties, for example in terms of depth and redshift range. In Figures
\ref{fig-lir_hist} and  \ref{fig-zed_hist} we compare the distributions of
\LIR\ and redshifts between the 3 surveys discussed above.  The range of \LIR\
between the three surveys is quite similar, with most galaxies in the interval
$43 <$ log \LIR\ $< 45$ erg/s, with the distribution peaking at log \LIR\ $\sim$ 44 erg/s.
The range for Herschel is somewhat narrower than AKARI or IRAS; the highest
luminosity galaxies are rare and therefore not expected to be frequent in small
area surveys.  Despite its enhanced sensitivity, Herschel also has fewer of
the lowest luminosity sources.  These can only be detected by AKARI and IRAS
at very low redshifts, the effective volume for which is again small for
the Herschel  Stripe 82 sample (particularly given the imposition of a 
low $z$ cut-off for the Herschel sample).

Figure  \ref{fig-zed_hist} shows that the Herschel detected sources are
biased towards considerably higher redshifts, on average, than AKARI and
IRAS.  Indeed, the redshift distribution of Herschel (whose median value
is 0.08) is closer to that of the complete SDSS Main Galaxy spectroscopic
sample (whose median redshift
is 0.10), whereas AKARI and IRAS are strongly biased towards relatively nearby
galaxies (median redshifts of 0.042 and 0.055 respectively).  The combination of
similar \LIR\ distributions, but different redshift ranges, indicates that the
AKARI and IRAS samples are likely biased towards galaxies with relatively
high SFRs.

\begin{figure}
\centering
\includegraphics[width=7.5cm,angle=0]{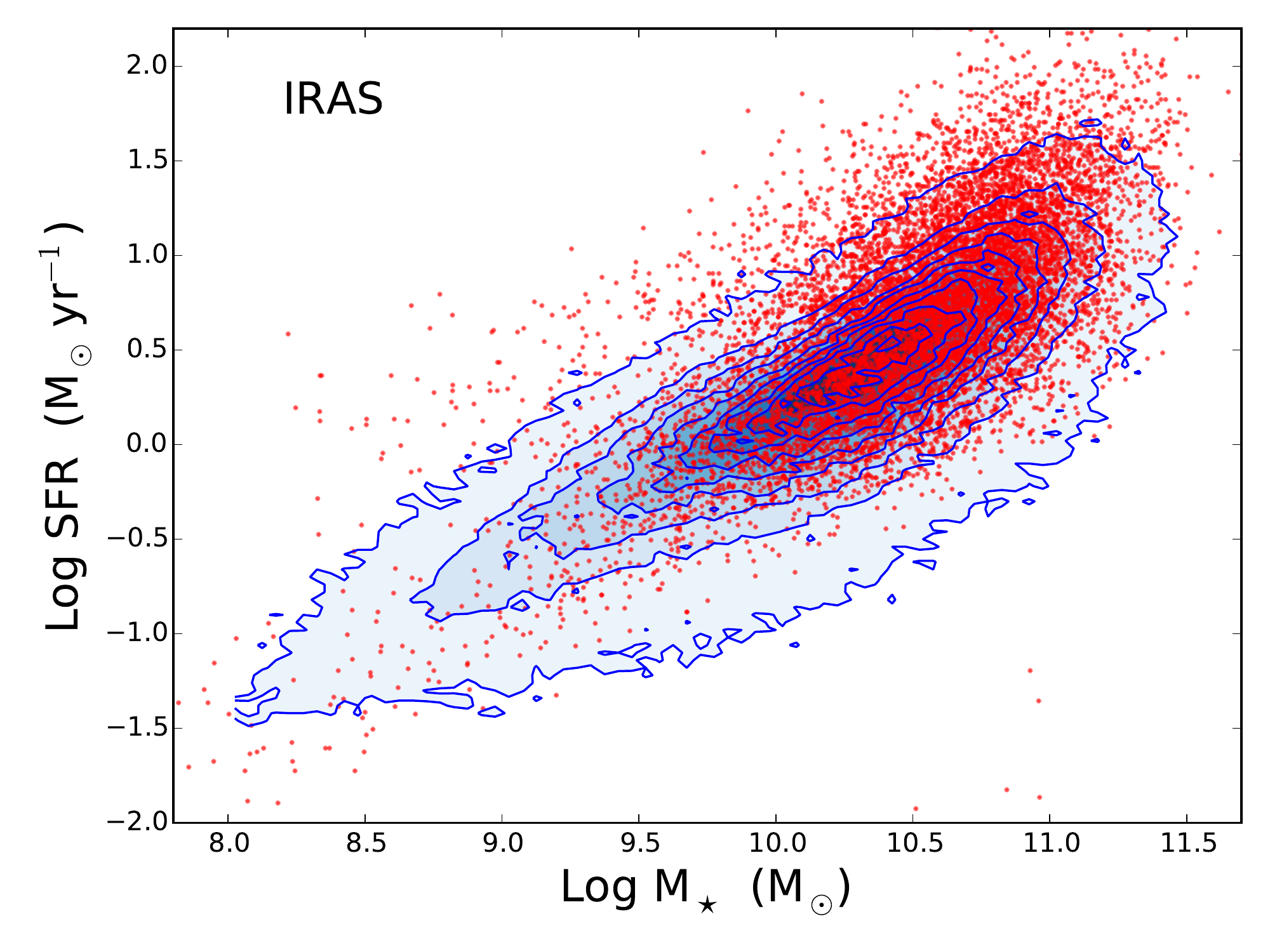}
\includegraphics[width=7.5cm,angle=0]{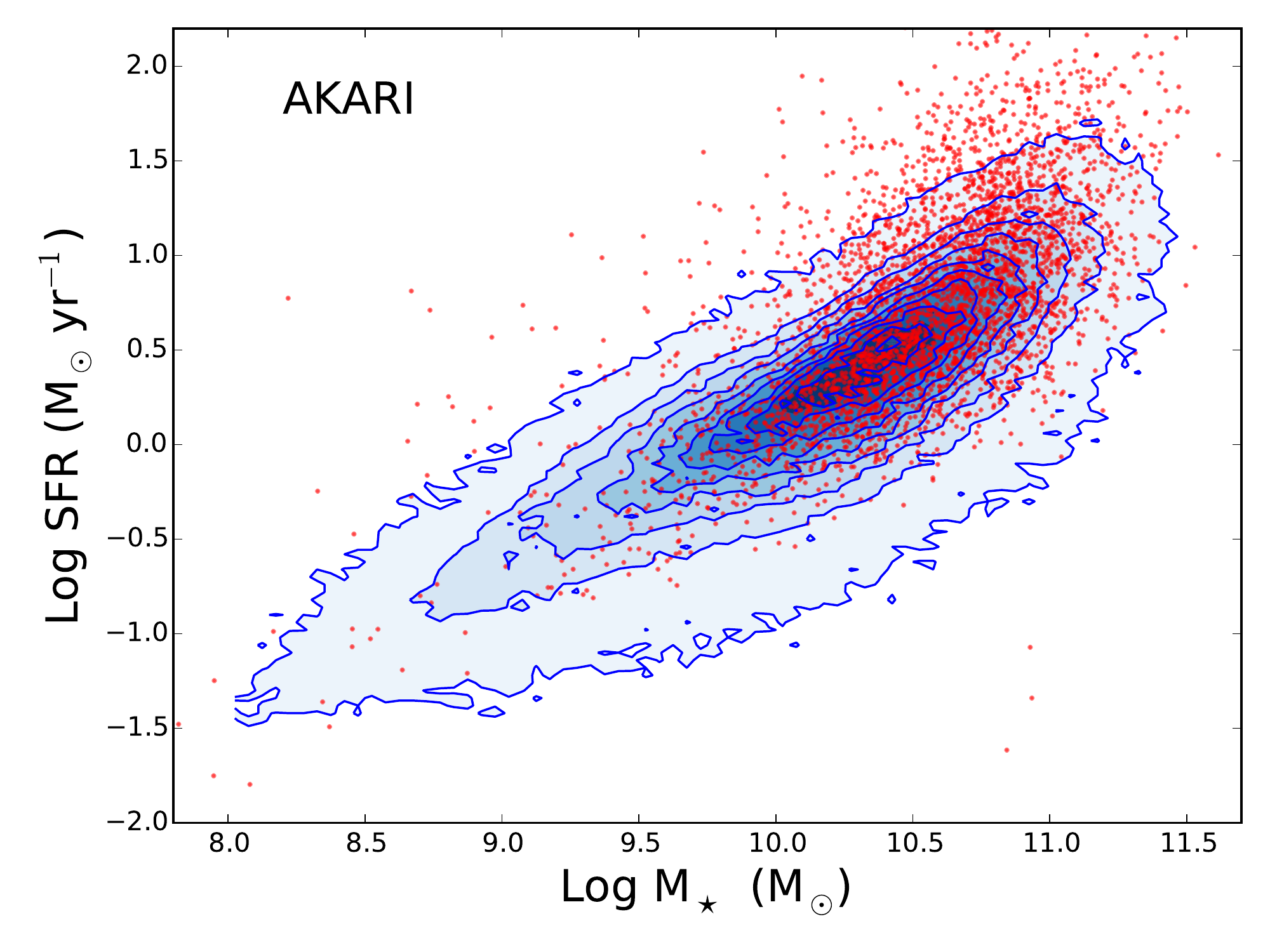}
\includegraphics[width=7.5cm,angle=0]{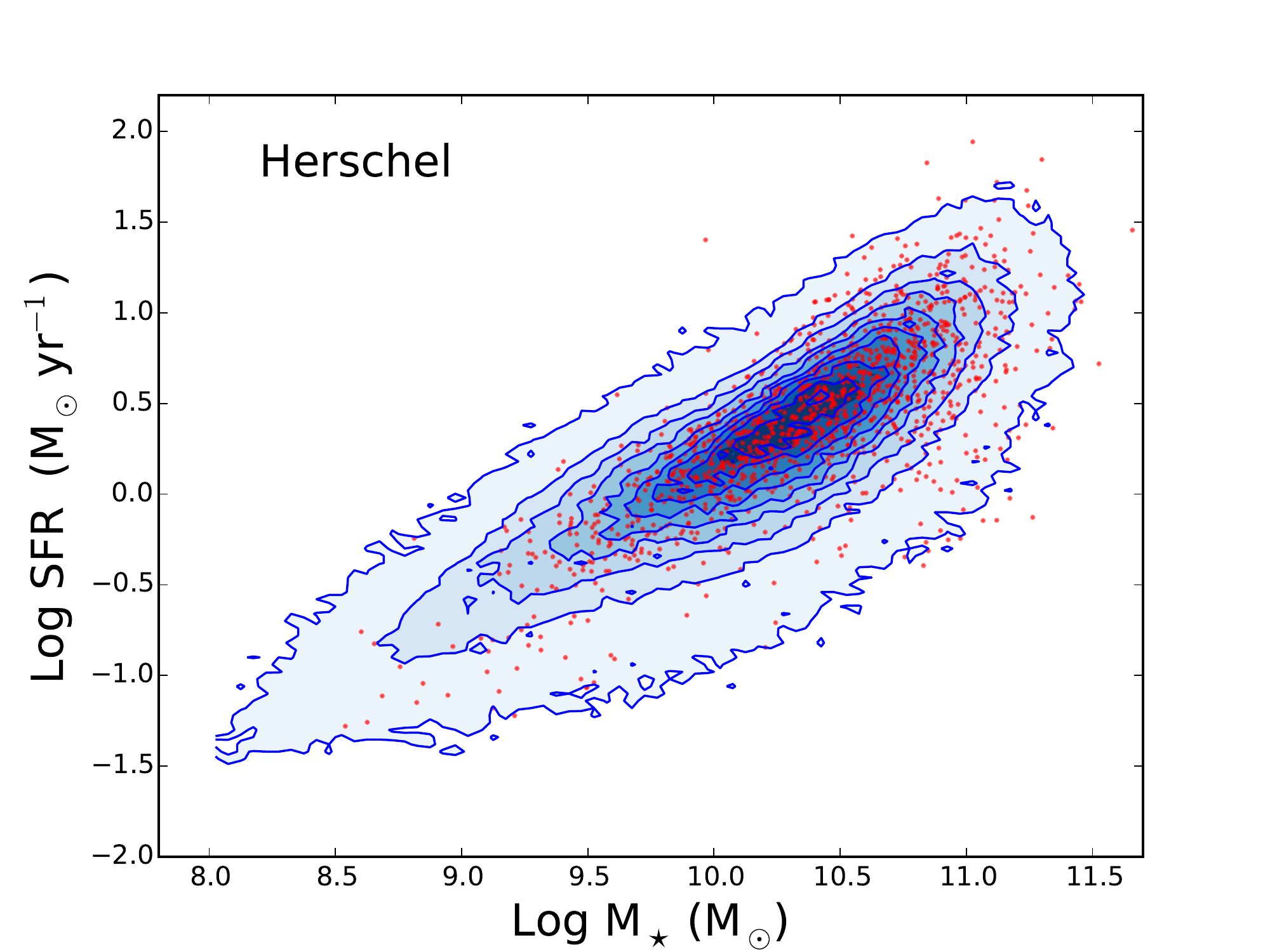}
\caption{The star-forming main sequence  for SDSS galaxies (filled
  contours in each panel) and SDSS matched  galaxies
  in the IRAS (top panel), AKARI (middle panel) and Herschel Stripe 82 (lower panel)
  datasets.  The SFRs for the SDSS star-forming galaxies are derived from
  aperture corrected spectroscopic values for galaxies classified as star-forming
  using the Kauffmann et al. (2003) diagnostic.  The SFRs for IR-detected galaxies
  are derived from the infra-red luminosities using Eqn. \ref{eqn-sfr} and shown
  as small points in each panel.}
\label{fig-ms}
\end{figure}

In Figure \ref{fig-ms} we show the total SFRs 
compared to the stellar masses, the so-called star-forming main sequence.
Star forming galaxies in the SDSS are shown in all three panels as
blue filled contours for reference, where
their SFRs are the aperture corrected values derived from the SDSS spectra.
Data points indicate the SFRs derived from IR luminosities
for each of the three IR surveys, computed using
Eqn. \ref{eqn-sfr}. Figure \ref{fig-ms} confirms that both AKARI and IRAS detected
galaxies are biased towards galaxies that have relatively high SFRs for their
stellar mass, whereas galaxies in the Herschel Stripe 82 sample are more typical
of the star-forming population as a whole.  There is also a significant population
of Herschel detected galaxies at relatively high masses that are below the main sequence.
Many of these are AGN dominated galaxies (using the diagnostic of Kauffmann et al.
2003); the results of Fig.  \ref{fig-ms}
are therefore consistent with previous results that low redshift 
optically selected AGN in the
SDSS can lie either on the main sequence or exhibit lower SFRs for their 
mass than star-forming galaxies (e.g. Salim et al. 2007; Gurkan et al. 2015)
 
In order to quantify the deviations from the
main sequence we define a SFR offset, that accounts not just for the mass
dependence of the main sequence but also redshift and local density.  For
every IR detected galaxy, we assemble a set of matched comparison galaxies
from the SDSS star forming sample (no IR detection required).
The comparison galaxies are required to have the same stellar
mass, redshift and local density as the IR detected galaxy, within
some tolerance.  Local density is defined as

\begin{equation}
\Sigma_n = \frac{n}{\pi d_n^2},
\end{equation}

where $d_n$ is the projected distance in Mpc to the $n^{th}$ nearest
neighbour within $\pm$1000 \kms.  Normalized densities, $\delta_n$,
are computed relative to the median $\Sigma_n$ within a redshift slice
$\pm$ 0.01.  In this study we adopt $n=5$ (although we note that the
results of our work do not rely on the choice of $n$).

The baseline tolerance used for matching is 0.1 dex in stellar mass, 0.005
in redshift and 0.1 dex in $\delta_5$.  We require at least 5 comparison
galaxies for each IR detected galaxy; if this is not achieved the mass, redshift
and local density tolerances
are grown in further increments of 0.1 dex, 0.005 and 0.1 dex respectively,
until the minimum size criterion of 5 is met.  However, there are typically
several tens of comparisons for each IR detected galaxy without the
need to extend the
tolerances.  The median log SFR of the matched comparison sample is subtracted
from the log SFR of the IR detected galaxy in order to determine the SFR
offset, $\Delta$ SFR, from the main sequence.

\begin{figure}
\centering
\includegraphics[width=8cm,angle=0]{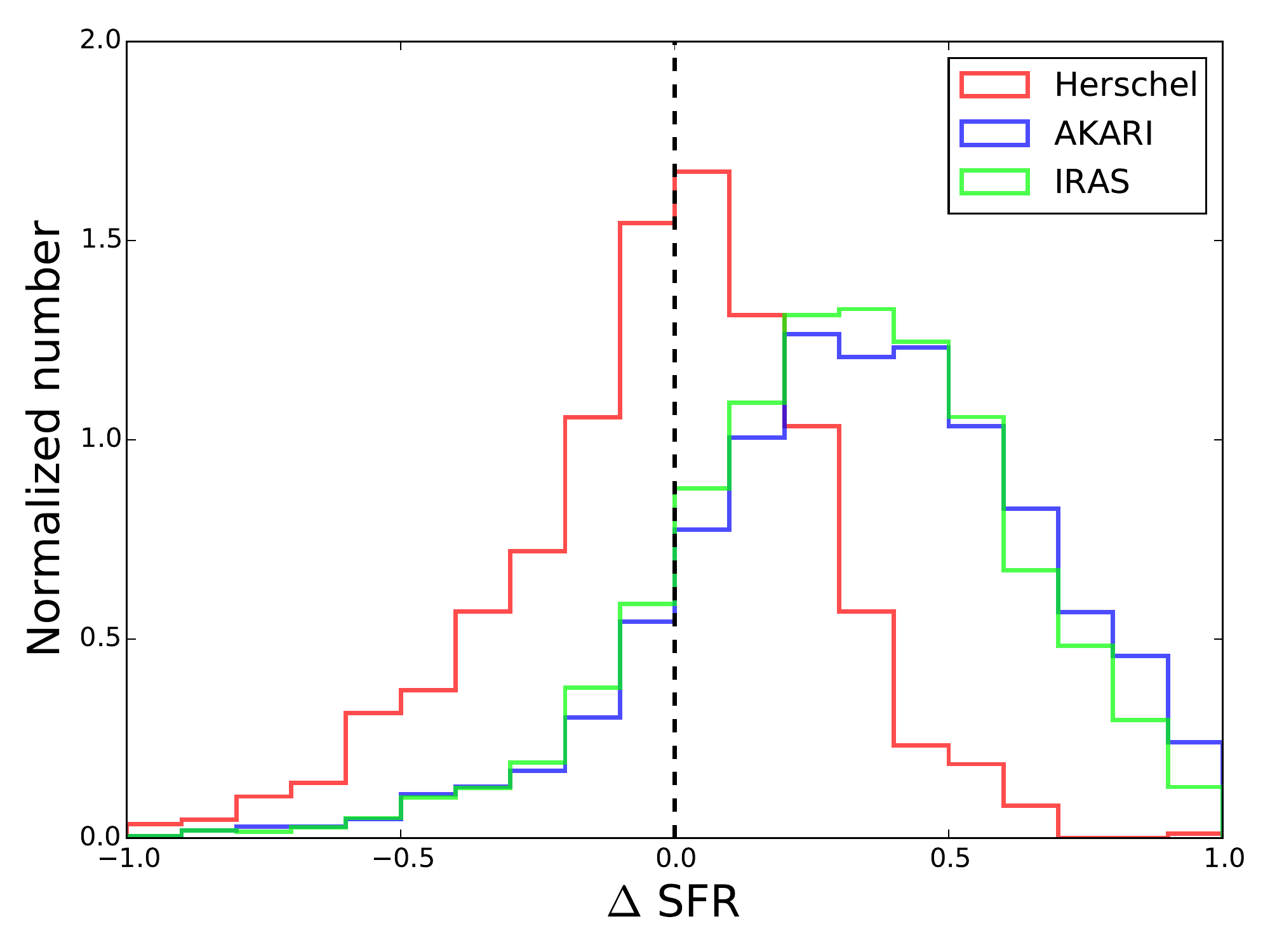}
\caption{The offset of IR detected galaxies from the star-forming main sequence
  defined by star-forming galaxies in the SDSS with the same stellar mass, redshift
  and local density.  The Herschel Stripe 82 sample (red histogram) has a median
  offset ($\Delta$ SFR = log SFR(IR detected galaxy) - log SFR(comparison median)) of
  approximately zero, implying that the sample is a good representation of normal star-forming
  galaxies.  However, the AKARI (blue) and IRAS (green) samples are skewed towards
  positive SFR offsets; on average, the galaxies detected in these two surveys
have SFRs higher by a factor of 2--3 compared to normal star forming galaxies.}
\label{fig-dsfr}
\end{figure}

In Figure \ref{fig-dsfr} we quantify the visual impression of elevated SFRs
in the IRAS and AKARI samples  from Fig. \ref{fig-ms} by plotting the
distribution of $\Delta$ SFR (the logarithmic SFR offset) for the three IR samples.
The Herschel Stripe 82 sample (red histogram) has a median offset of approximately
zero, demonstrating that these galaxies are generally drawn from the normal
star-forming main sequence.  In contrast, both the AKARI and IRAS samples
(blue and green histograms respectively) are strongly skewed to positive
$\Delta$ SFR.  The galaxies detected in these two surveys are typically
forming stars at rates 2--3 times higher than their comparison galaxies (i.e.
at fixed stellar mass, redshift and environment).  AKARI and
IRAS are therefore not suitable training sets for the ANN, since training
with biased data will yield a network that similarly predicts SFRs that
are relatively high.  Nonetheless, AKARI and IRAS can still be used
as validation sets.  Indeed, it is a good test of the network's reliability
if it is able to accurately predict the characteristics of a sample
whose properties contrast with its training set.

 In summary of this section:  we have homogeneously determined the IR luminosities
 of 14,255 galaxies in IRAS all sky survey, 6957 galaxies in the AKARI all sky survey
 and 1349 galaxies in the Herschel Stripe 82 survey that have matches in the
 SDSS spectroscopic survey.  The IRAS photometry yields \LIR\ values that are
 offset by 0.15 dex from what we expect based on their SDSS SFRs; a similar offset
 is seen in the comparison of \LIR\ for galaxies in common between IRAS and either
 AKARI or Herschel.  We also show that AKARI and IRAS are biased towards relatively
 high SFR galaxies, whereas the Herschel Stripe 82 sample is a good representation
 of star-forming galaxies at its median redshift of $z \sim 0.08$.  For these reasons,
 the Herschel sample is our preferred training sample.  Although it is the smallest
 of the three catalogs, its population is unbiased and its photometry is accurate
 (see also Rosario et al. 2015).
 
\section{The artificial neural network and determination of infra-red luminosities}\label{sec-ann}

Although many varieties of ANN exist, all of which can be configured, trained and optimized
to solve
various data challenges, the basic functionality of these techniques
is similar.
Based on a collection of input (training) data, the network derives a set of optimal weights
and biases that best describes the relation to the target data.  In this way, an ANN
can be used to `predict' target values for an independent dataset from its input
variables.  There is a vast literature available on the topic of ANN (e.g. Chen et al
1991; Battiti 1992; De Jesus \& Hagan 2007), including
its application to astronomical data problems (Andreon et al. 2000; Ball et al. 2004;
Teimoorinia 2012; Gonzalez-Martin et al. 2014; Teimoorinia \& Ellison 2014).  
Here we provide a high level overview of
the statistical techniques adopted in the current work, referring the reader to
technical papers (e.g. Bishop 2007) for more details. 

\subsection{Training parameters}

The ultimate objective of the current work is to train a network that will
predict the \LIR\ for a large fraction of the galaxies in the SDSS, as
justified in the Introduction.  The starting point of this work is
therefore the identification of a sizable, high quality
and homogeneous dataset for which we have a wide variety of galaxy
parameters from SDSS available for training, and matched IR luminosities
to act as the target data.  The procedure will then be to train the
network to recognize the linkages between SDSS parameters and \LIR,
and then apply the trained network to predict the \LIR\ for the
rest of the SDSS.

Fortunately, there are numerous public
catalogs of SDSS galaxy properties, both measured and derived,
from both spectroscopic and photometric products (e.g. Kauffmann et al. 2003;
Brinchmann et al. 2004; Simard et al. 2011; Mendel et al. 2014).  From this
extensive vault of data, we experimented with various combinations of training
parameters.  Since the infra-red luminosity is largely determined by the
galaxy's SFR, our first attempts at network training focussed on the combination
of stellar mass, the fluxes of various emission lines and the strength
of the 4000 \AA\ break.  Emission line fluxes are taken from the public MPA/JHU
catalogs, that have been corrected for underlying stellar absorption and Galactic
extinction.  We apply further corrections to the line fluxes for internal extinction
as described in Scudder et al. (2012) by adopting a Small Magellanic
Cloud (SMC) extinction curve (Pei 1992), before converting the
fluxes to luminosities.

In recognition
of the limiting fibre aperture, we included both total and fibre stellar masses
in the training set, as well as the explicit covering fraction of the $r$-band
light.  It was found that the network's performance could be further improved
with the provision of colours and absolute magnitudes.  We also experimented
with various morphological information (based on the bulge-disc
decompositions of Simard et al. 2011), but no significant improvement was
obtained.  Our final set of training parameters is presented in Table
\ref{tab-input}.  
There are 8602, 3153 and 1136 galaxies in the IRAS, AKARI and Herschel Stripe
82 sample, respectively, for which all 23 of these parameters are available,
where the majority of the IRAS and AKARI galaxies have been excluded due to
their bright optical magnitudes which preclude them from many of the SDSS
spectroscopic samples (e.g. Simard et al. 2011; Mendel et al. 2014).

\begin{table}
\begin{center}
\caption{The  parameters  used in work as input data for training and validation sets.}
\begin{tabular}{l}
\hline
M$_{\star, fibre}$: Fibre stellar mass\\
M$_{\star}$: Total stellar mass\\
Corrected luminosity of [OII] $\lambda$ 3727 \AA\ emission line \\
Corrected luminosity of [OII] $\lambda$ 3729 \AA\ emission line \\
Corrected luminosity of [OIII] $\lambda$ 4959 \AA\ emission line  \\
Corrected luminosity of [OIII] $\lambda$ 5007 \AA\ emission line  \\
Corrected luminosity of H$\alpha$ $\lambda$ 6563 \AA\ emission line  \\
Corrected luminosity of H$\beta$ $\lambda$ 4861 \AA\ emission line  \\
Corrected luminosity of [NII] $\lambda$ 6582 \AA\ emission line  \\
Corrected luminosity of [SII] $\lambda$ 6717 \AA\ emission line  \\
Corrected luminosity of [SII] $\lambda$ 6731 \AA\ emission line  \\
$z$: Redshift  \\
D$_{4000}$: 4000 \AA\ break\\
$r$-band covering fraction  \\
M$_u$: absolute $u$-band magnitude \\
M$_g$:  absolute $g$-band magnitude  \\
M$_r$:  absolute $r$-band magnitude  \\
M$_i$:  absolute $i$-band magnitude  \\
M$_z$:  absolute $z$-band magnitude  \\
u-g observed colour   \\
g-r observed colour  \\
r-i observed colour \\
i-z observed colour  \\
\hline
\end{tabular}
\label{tab-input}
\end{center}
\end{table}

\subsection{Training algorithms and error estimation}\label{error_sec}

After configuring the network for the input data, the main task of the ANN is to
determine the appropriate weights and biases that best connect them to the target
data.  This is an iterative process, and can be achieved with different algorithms
that are variously suited to different data situations.  In the current work
we derive the weights and biases according to the Levenberg-Marquardt optimization
method (Marquardt 1963).
This common technique minimizes a combination of squared errors and
weights (Eq. \ref{eq-br}), and then determines the correct combination.  An adaptation
to the basic  Levenberg-Marquardt optimization, that may help to 
produce a network that generalizes well, is the addition of a Bayesian regularization term.
The addition of the Bayesian regulator is particularly useful if the training set is not
very large  (e.g., Neal, 1996).  Specifically, we define the error function of the
weights (\textbf{W}) as

\begin{equation}
\rm{Err(\textbf{W})=\sum\limits_{i=1}^N  [Y(\textbf{X}_i,\textbf{W})-T_i]^2 + \lambda \textbf{W}^T\textbf{W} }
\label{eq-br}
\end{equation}

in which Y and T are the output of the network and the target values, and \textbf{X} are the
input data (in this case, the variables listed in Table \ref{tab-input}) for N galaxies.  The Bayesian
regularization approach is important in the control of the so-called `over-fitting problem' (Bishop 2007),  which
can occur when we use a powerful network (i.e., a network with large number of neurons) or the number of
iterations in the training step (for updating the weights) is large.  Over-fitting is also a pitfall
associated with having a small training set.  The Bayesian regularization mitigates the potential
for over-fitting by adding a term to the squared error function (the first term in equation \ref{eq-br});
this is sometimes referred to in ANN models as weight decay.   In Eq. \ref{eq-br}, the coefficient $\lambda$
governs the relative importance of the regularization term compared with the error function of Eq. \ref{eq-br}.

\begin{figure}
\centering
\includegraphics[width=8cm,height=6cm,angle=0]{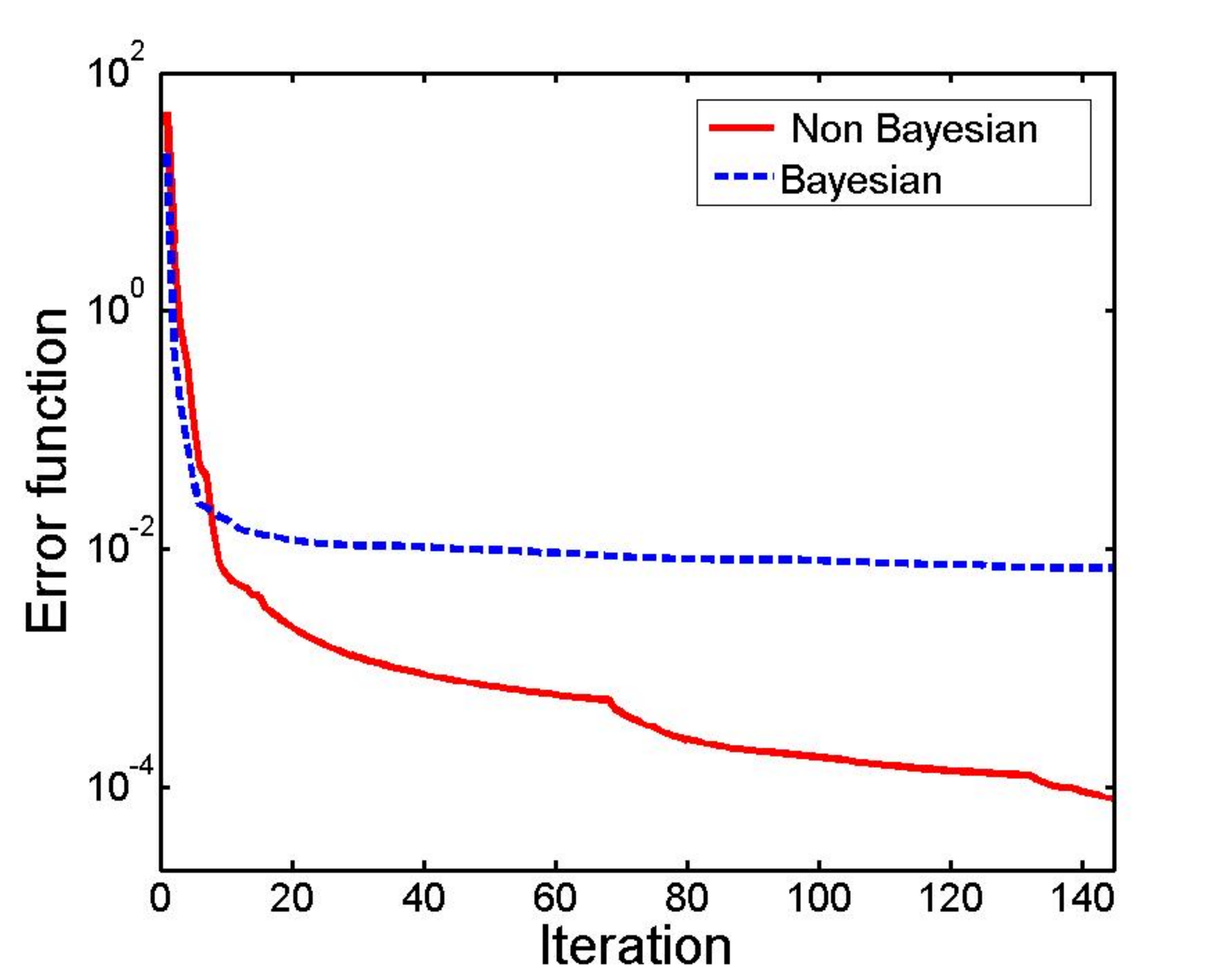}
\caption{The performance of Levenberg-Marquardt optimization method  without (solid line, the first term in equation \ref{eq-br})
  and with a Bayesian regulator term (dashed line, addition of the second  term in equation \ref{eq-br}) for a network using 50 neurons.
  Without the Bayesian regulator, there is a continued improvement as the number of iterations increases, indicative of an over-fitting problem.
The addition of the Bayesian regularization shows convergence.}
\label{fig-br-lm}
\end{figure}

In many situations, the Levenberg-Marquardt algorithm can be successfully used to minimize only the first term of
Eq. \ref{eq-br}, removing the need for the Bayesian regularization. To demonstrate the differences between the two
methods (with and without Bayesian regularization), we use a powerful network (N$_{neuron}$=50) and  also use a
large number of iterations to update the weights with the Herschel data used as the training set. The result can
be seen in Figure \ref{fig-br-lm}. In the Bayesian regularization approach, no significant improvement in the network
performance can be seen after $\sim$20 iterations (the blue dashed line). If we  minimize  only the first term of
Eq. \ref{eq-br} the result will show a better performance for higher iterations (red dashed line) but it has tendency
to be over-fitted, so it is not suitable for generalization.  For a more secure result, it is desirable to
decrease the number of neurons, if the reduction does not decrease the performance of the network.
In this work, we have found that decreasing the number of neurons to $\sim 15$ still yields an almost identical
output as the more powerful 50 neuron network shown by the blue line in Figure  \ref{fig-br-lm}, and
with no absolute  improvement after $\sim 50$ iterations.

The network is now fully trained and ready to be applied to a target dataset.  The ultimate
target dataset (we will investigate some validation sets in the following section) is the
complete SDSS, for which all 23 parameters listed in Table \ref{tab-input} are available (331,926 galaxies).
However, although we have shown that the Herschel Stripe 82 sample contains typical star forming
galaxies, the training set may not be representative of \textit{all} galaxies.
Although the network can extrapolate well behaved correlations, it may
not yield accurate predictions in cases where the target data exhibit different properties
to the training set.  We therefore develop a technique to identify galaxies with less
secure \LIR\ predictions.

First, to avoid any bias and to obtain a statistical result, we repeat the training procedure
25 times and select the best 20 trained networks. This step is necessary because sometimes the initialization
of the parameters can yield a solution with very high scatter, indicating that the network was not well trained, e.g. it has converged on a local, rather than global minimum in parameter space.
The output of the  best 20 trained networks show small differences due to the different initializations.
The average value of the predicted \LIR\ for a given galaxy is computed as
$\textbf{(W$_k)$}^T \textbf{X}$ for k=1 to 20, where \textbf{W$_k$} are the weights obtained from the
20 best trained networks.   Since \textbf{W$_k$} have been determined for the Herschel input parameter space,
for those galaxies in the SDSS sample that have very different input data, \textbf{X} , the scatter will be very
large.   On the other hand, for the galaxies that have input parameter space similar to Herschel's,
the variations should be small,  because the weights and biases are tailored for this input data.
We adopt the mean \LIR\ value of the 20 best trained networks, and assign the associated `error'
($\sigma_{\rm ANN}$) as the scatter amongst the network outputs.  This error does not encapsulate the uncertainties
in the calibration itself, nor those associated with the input data.  Rather, it provides
a confidence indicator for the network predictions, which we will later use
to identify robust samples.

\section{Network validation}\label{sec-valid}

\begin{figure}
\centering
\includegraphics[width=7.5cm,height=6cm,angle=0]{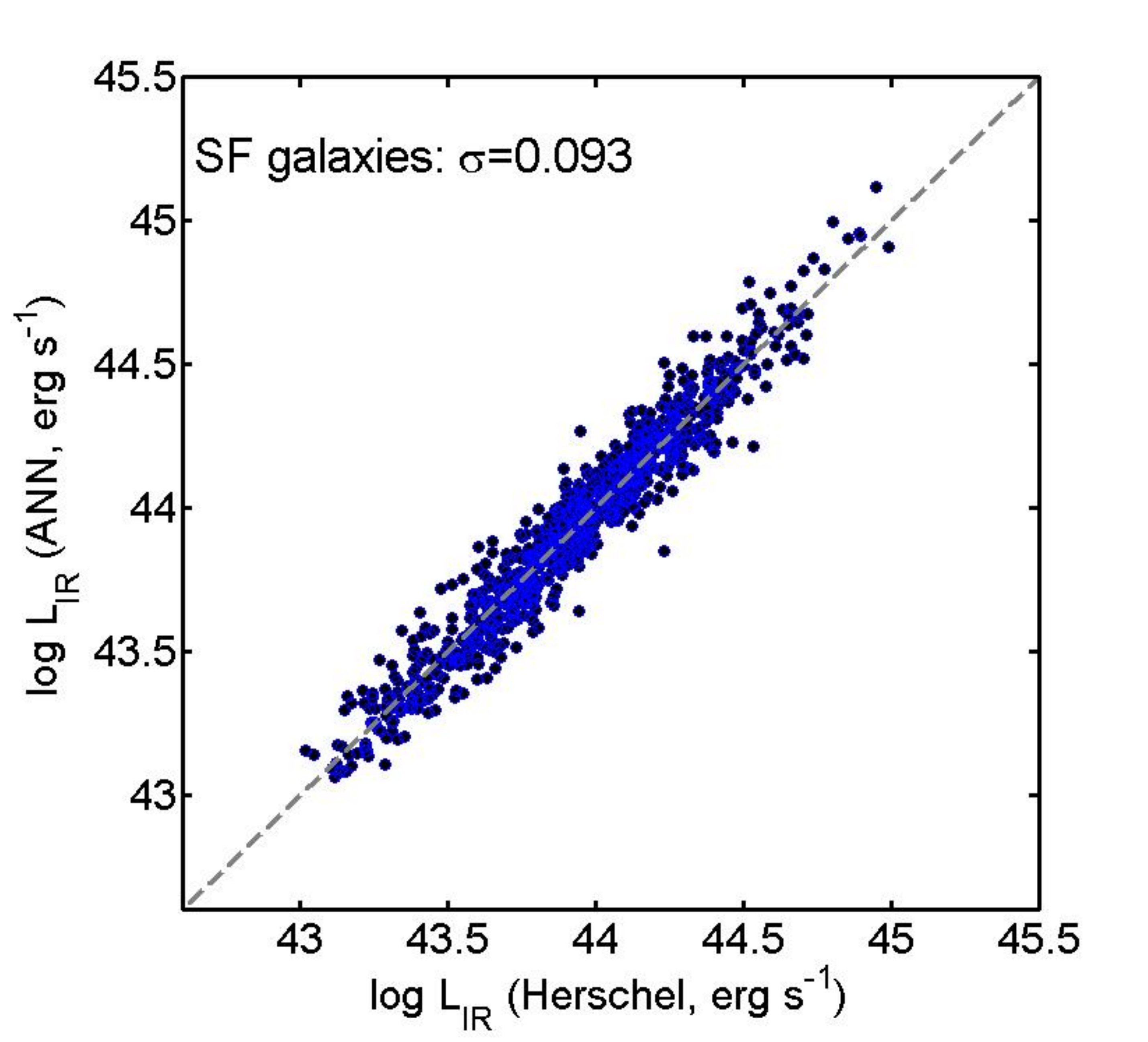}
\includegraphics[width=7.5cm,height=6cm,angle=0]{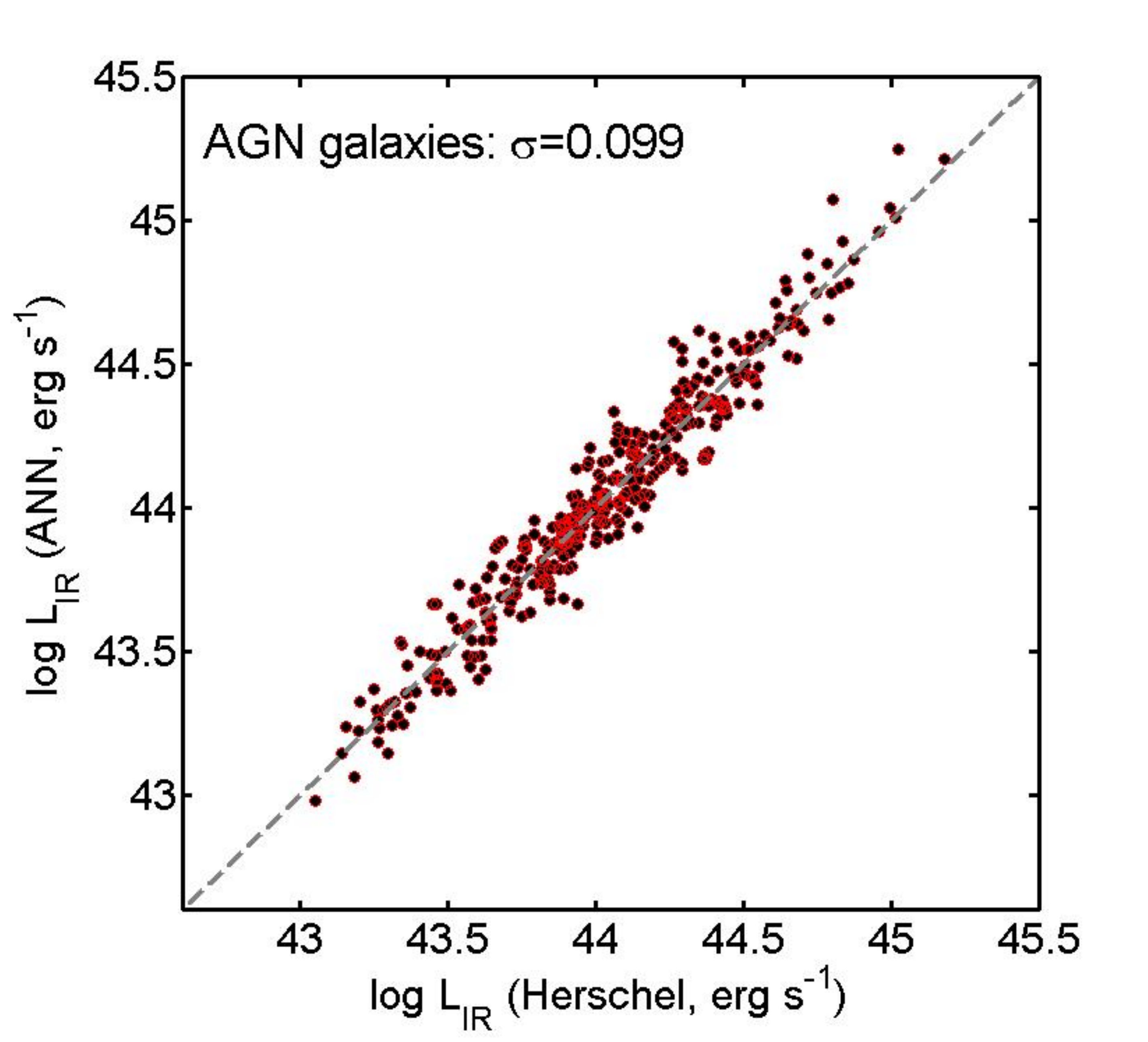}
\caption{Self validation of the ANN with Herschel Stripe 82 data.  The
  ANN-predicted \LIR\ is compared with observed infra-red luminosities
  for the training sample.  The top panel shows the results for galaxies
  classified as star-forming (Kauffmann et al. 2003) and the lower panel
for AGN.  The network is well trained for both galaxy classifications.}
\label{fig-pred-s82}
\end{figure}

\subsection{Self validation}

The first validation check involves an internal (self) validation with the training dataset.
If the network is well trained, the performance of the self-validation should be good.
In Figure \ref{fig-pred-s82} we show the predicted \LIR\ from the ANN compared to
the Stripe 82 luminosities.  We assess the performance of the ANN separately for
star-forming galaxies and those with an AGN.  Throughout this paper
we distinguish AGN using the Kauffmann et al. (2003) classification,
with a S/N$>$1 requirement in all 4 necessary lines.
In the top panel of Figure \ref{fig-pred-s82},
we show the results for the star-forming galaxies; the performance
is excellent, with a scatter between the observed and predicted \LIR\ of only
0.093 dex.  Such a small scatter is expected for three reasons.  First, in the
self validation step we are testing on the data used to train the network, so
we know that the optimum weights and biases for this dataset are being applied.
Second, of the 23 input parameters used (Table \ref{tab-input}), several
(such as the emission line fluxes) are known to correlate very well with SFR,
and should hence be excellent indicators of \LIR.  Finally, the star-forming
sample represents the bulk of the training set (892/1136 galaxies).  In the
lower panel of  Figure \ref{fig-pred-s82} we show the performance of the AGN
dominated galaxies.  Again, the performance is excellent, with a scatter of
0.099 dex, indicating that the network is equally well trained and applicable
to galaxies with an active nucleus.  This is a non-trivial result: the emission
lines in AGN galaxies have contributions from more than one photo-ionizing source.
Therefore, in contrast to the star-forming galaxies, there is no \textit{a priori}
mapping between emission line fluxes, SFRs and \LIR.
Nonetheless, the ANN is able to combine the inputs of all the parameters, with
appropriate weights and biases, to determine a solution for \LIR\ that is
applicable even when the AGN `contamination' is present.

\subsection{Validation with independent datasets}

\begin{figure*}
\centering
\includegraphics[width=18cm,angle=0]{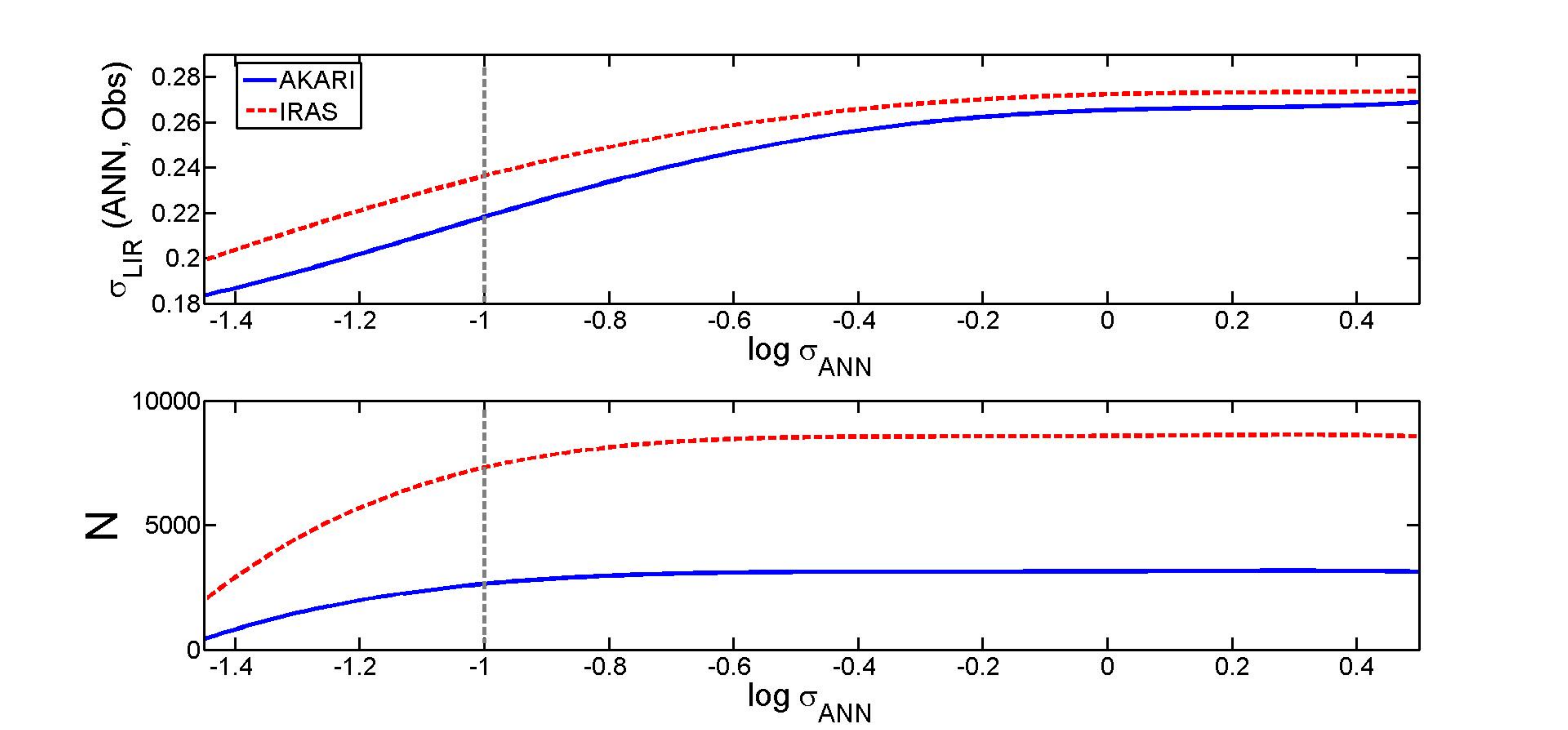}
\caption{The effect on sample size (lower panel) and scatter (upper panel) as a function
of $\sigma_{\rm ANN}$ threshold for the two independent validation sets (IRAS and AKARI).  
More stringent (smaller) thresholds in $\sigma_{\rm ANN}$
leads to a smaller scatter in the prediction of \LIR\ by the ANN, but reduces the
size of the sample.  A nominal requirement of $\sigma_{\rm ANN} < 0.1$ (vertical dashed lines)
is adopted for this work as a balance between sample size and scatter. }
\label{sig_cut}
\end{figure*}

\begin{figure}
\centering
\includegraphics[width=9cm,angle=0]{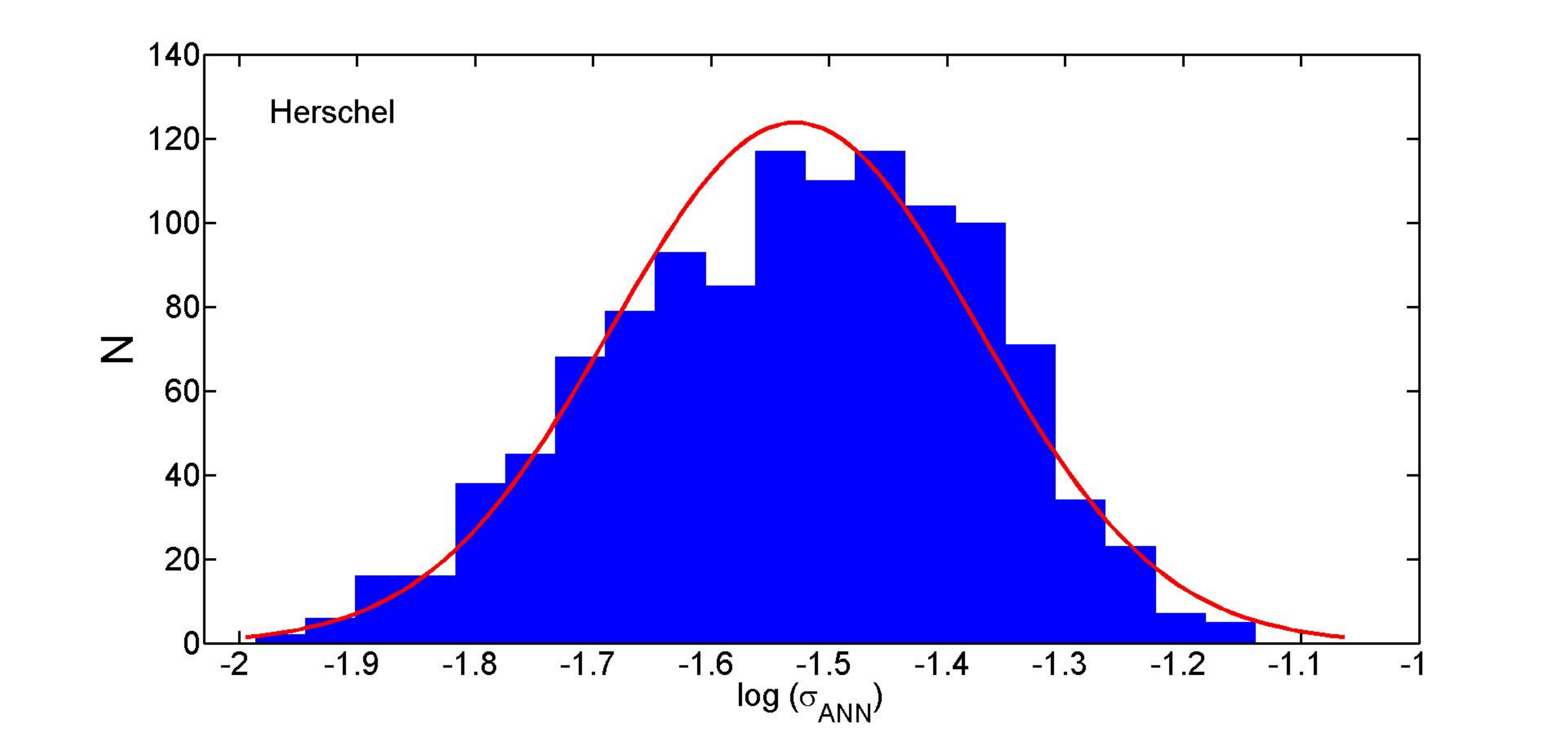}
\caption{The distribution of $\sigma_{\rm ANN}$ in the Herschel training set.  A Gaussian
fit to the distribution is shown in red. }
\label{sig_her}
\end{figure}

Whilst the results of the self-validation are re-assuring, the next critical step is to
validate the ANN on independent datasets; AKARI and IRAS are used for this purpose.
Before investigating in detail the performance of the independent validation, it is
useful to quantify the effect of cuts in $\sigma_{\rm ANN}$ on the two datasets.
We define $\Delta$ \LIR\ as the difference between the ANN predicted \LIR\ and
the measured value.
In Fig. \ref{sig_cut} we show the scatter ($\sigma_{LIR}$) 
in $\Delta$ \LIR\ for the AKARI and IRAS datasets, as a function of $\sigma_{\rm ANN}$
threshold.    The choice of where to put the error threshold
is a trade-off between the size of the sample for which \LIR\ is predicted, and
the likely scatter in those predictions, and different thresholds in $\sigma_{\rm ANN}$
will be appropriate for different applications.  In Fig. \ref{sig_her} the distribution of 
of $\sigma_{\rm ANN}$ for the Herschel training set is plotted, with
a Gaussian fit overlaid.  The tail of the Gaussian distribution extends to $\sigma_{\rm ANN}
\sim 0.1$ which we adopt as a working threshold for the remainder of this paper.

\subsubsection{AKARI validation}

\begin{figure}
\centering
\includegraphics[width=7.cm,height=5.5cm,angle=0]{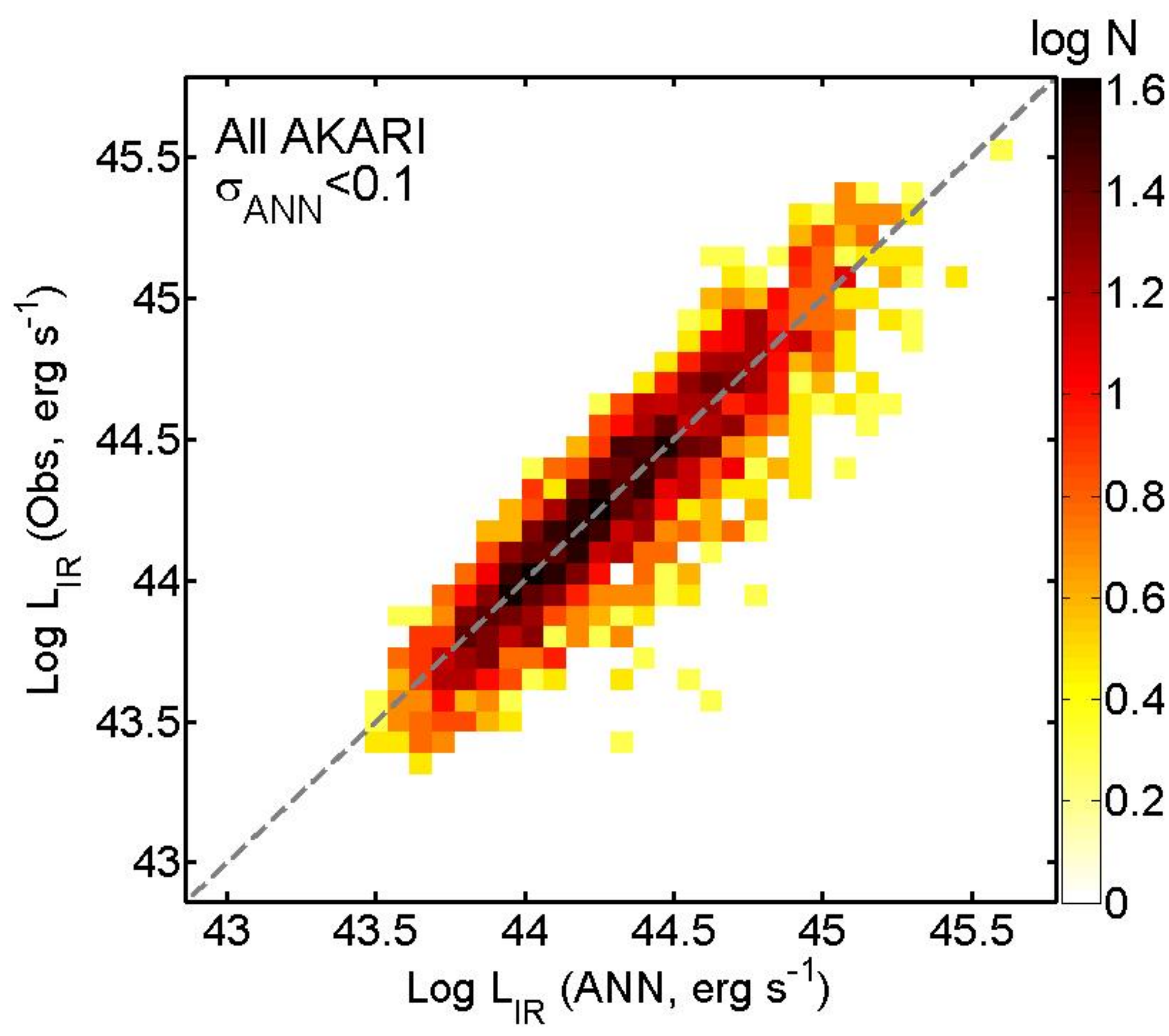}
\includegraphics[width=7.cm,height=5.5cm,angle=0]{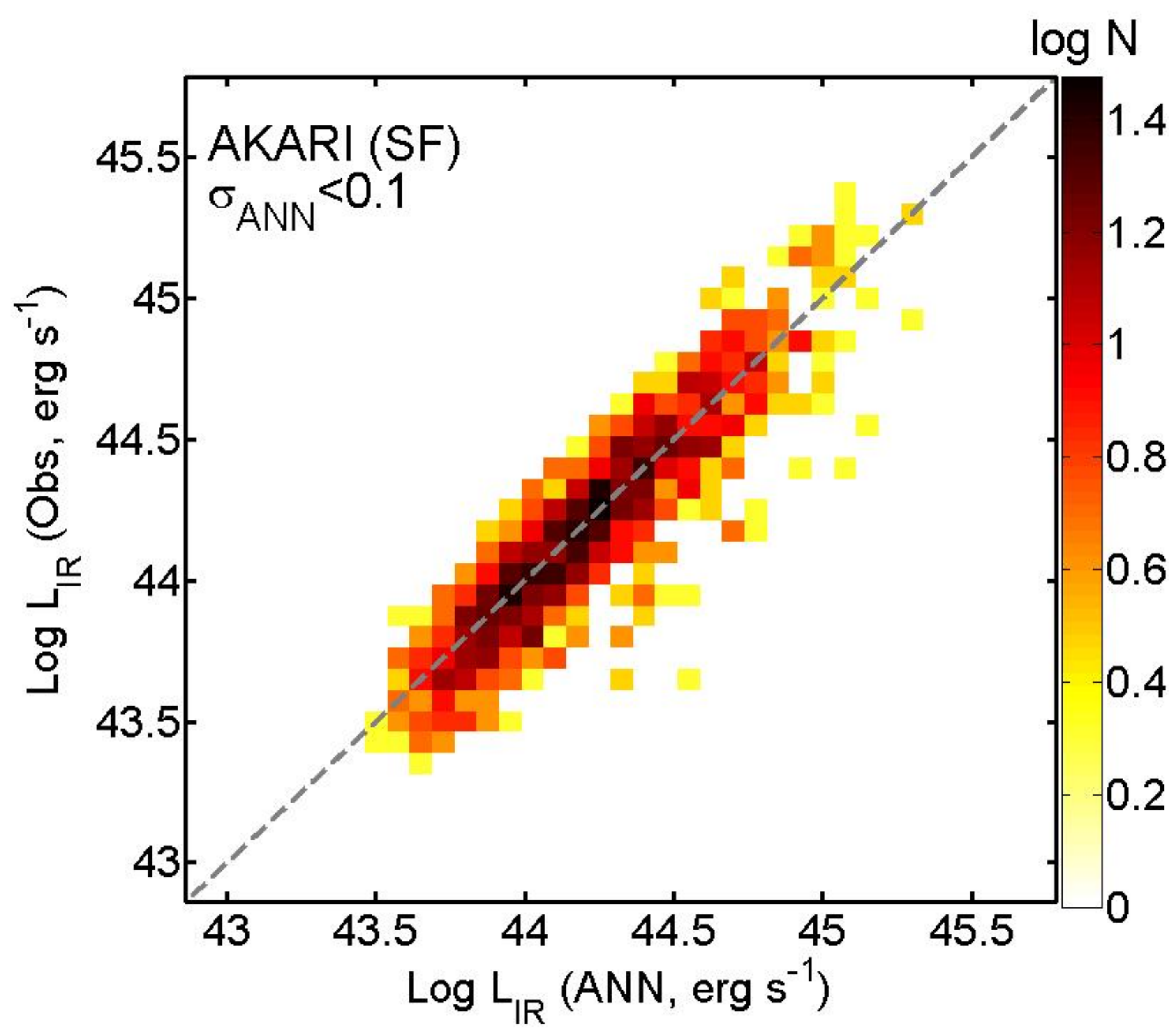}
\includegraphics[width=7.cm,height=5.5cm,angle=0]{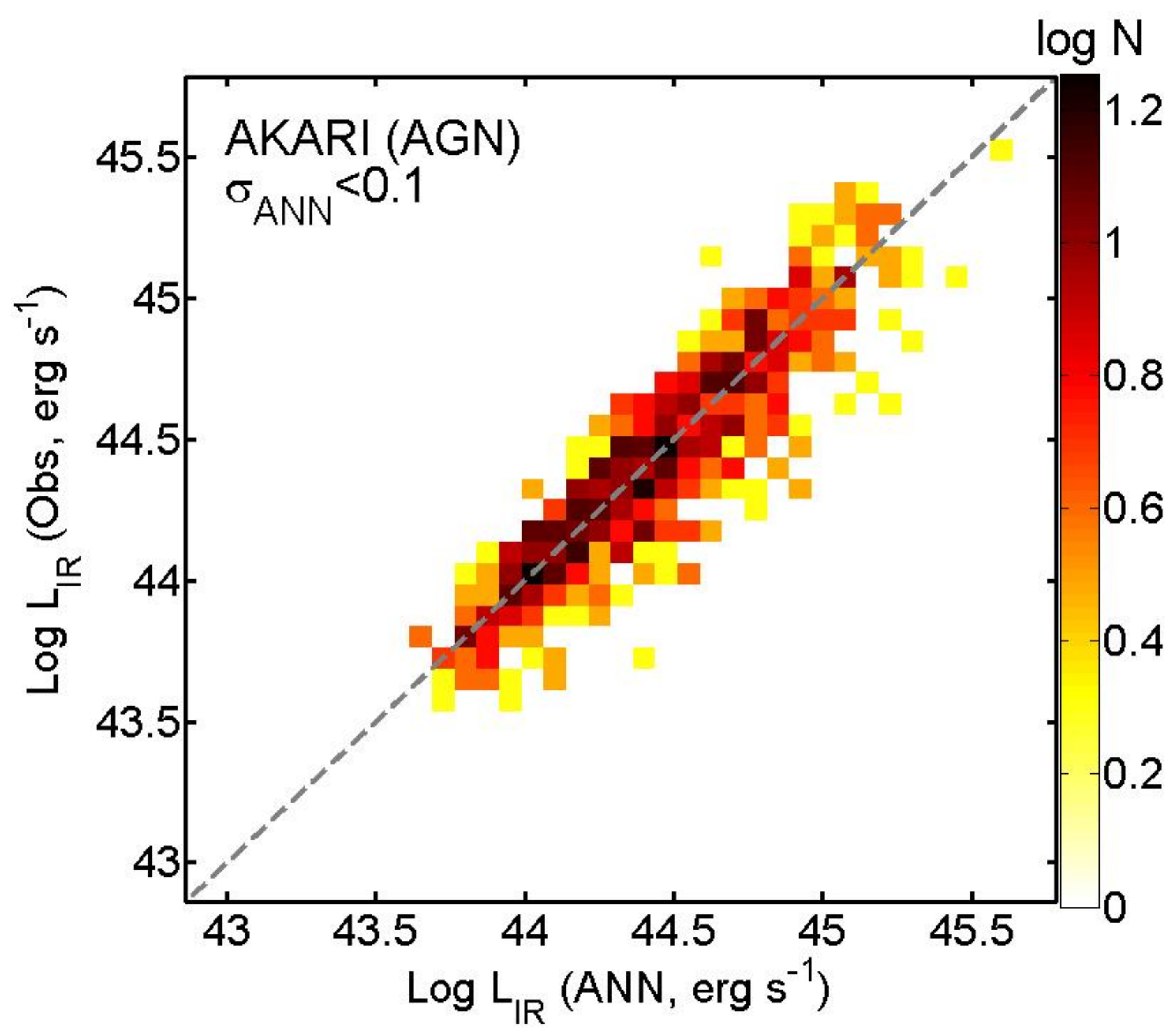}
\caption{A comparison between the observed and predicted \LIR\  for the AKARI sample;
all panels are colour coded by galaxy density and have an error cut of $\sigma_{\rm ANN} < 0.1$.
  The top panel shows all galaxies, and the middle and lower panels distinguish 
the star forming and AGN galaxies respectively.  The scatter for the star-forming
and AGN galaxies is comparable: 0.23 and 0.25 dex, respectively.}
\label{fig-akari}
\end{figure}

In the top panel of Figure \ref{fig-akari} we compare the ANN-predicted and AKARI \LIR\
for the 2649 galaxies whose ANN error is $\sigma_{\rm ANN}<$ 0.1 dex, with points colour-coded by
the density of galaxies in each cell.  Recall that $\sigma_{\rm ANN}$ is determined from
the distribution of \LIR\ over
20 different initializations of the training, and is a reflection of the similarity of
the input variables of a given test galaxy and the training set.  The scatter
in the comparison between measured AKARI IR luminosities and those predicted
by the ANN after the $\sigma_{\rm ANN}<$0.1 dex threshold has been imposed is 0.22 dex,
with a symmetric scatter.
Although this scatter is higher than for the
self validation (which is to be expected), there is no systematic offset between
the prediction and observed \LIR.  As noted in Section \ref{akari_sec},
at least part of the increase in scatter in the comparison with AKARI is likely
due to larger uncertainties in AKARI's photometry. Indeed, we experiment
with using AKARI as the initial training set and find that it has a much
higher internal scatter than the Herschel data.

In the middle and lower panels of  Figure \ref{fig-akari} we impose the same error
cut of 0.1 dex, but now separate the galaxies that are classified as star-forming or AGN.
The scatter in the two sub-samples is very comparable, 0.20 dex and 0.24 dex for
the star-forming and AGN classes, respectively,  demonstrating
that the single network can equally well be used to derive infra-red luminosities
independently of the emission line classification.

\subsubsection{IRAS validation}

Despite the presence of an offset between the IRAS photometry and that
of Herschel and AKARI, the IRAS dataset can still play a useful role
in the validation of our trained network, thanks to its very large
size. Recall that a 0.15 dex correction is applied to the IRAS
IR luminosities in order to account for this offset.
As was shown in the lower panel of Fig. \ref{fig-compare-cats-sfr},
once the offset is applied, there is a  good average agreement between the SDSS SFRs and
IRAS \LIR.

\begin{figure}
\centering
\includegraphics[width=7.cm,height=5.5cm,angle=0]{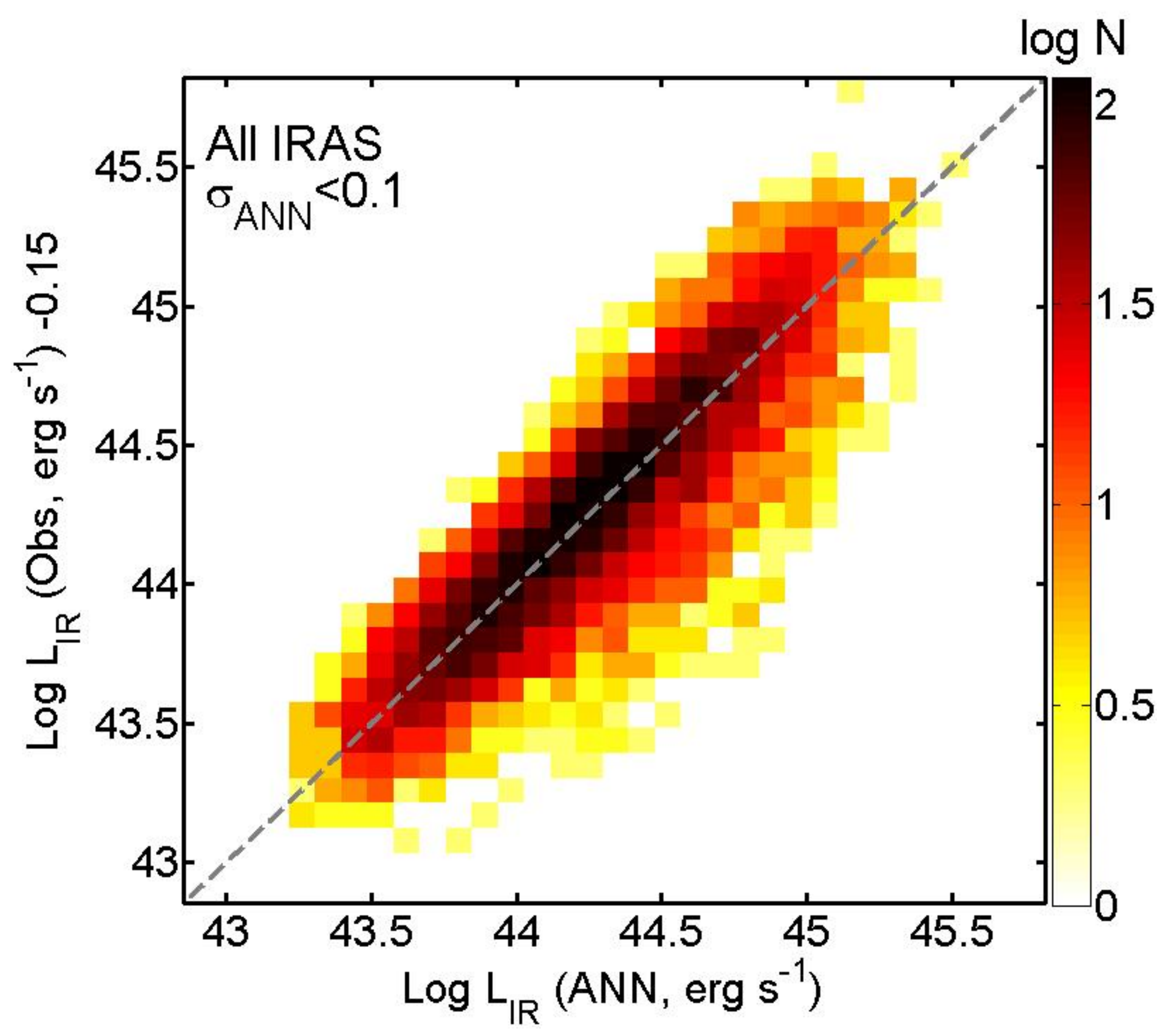}
\includegraphics[width=7.cm,height=5.5cm,angle=0]{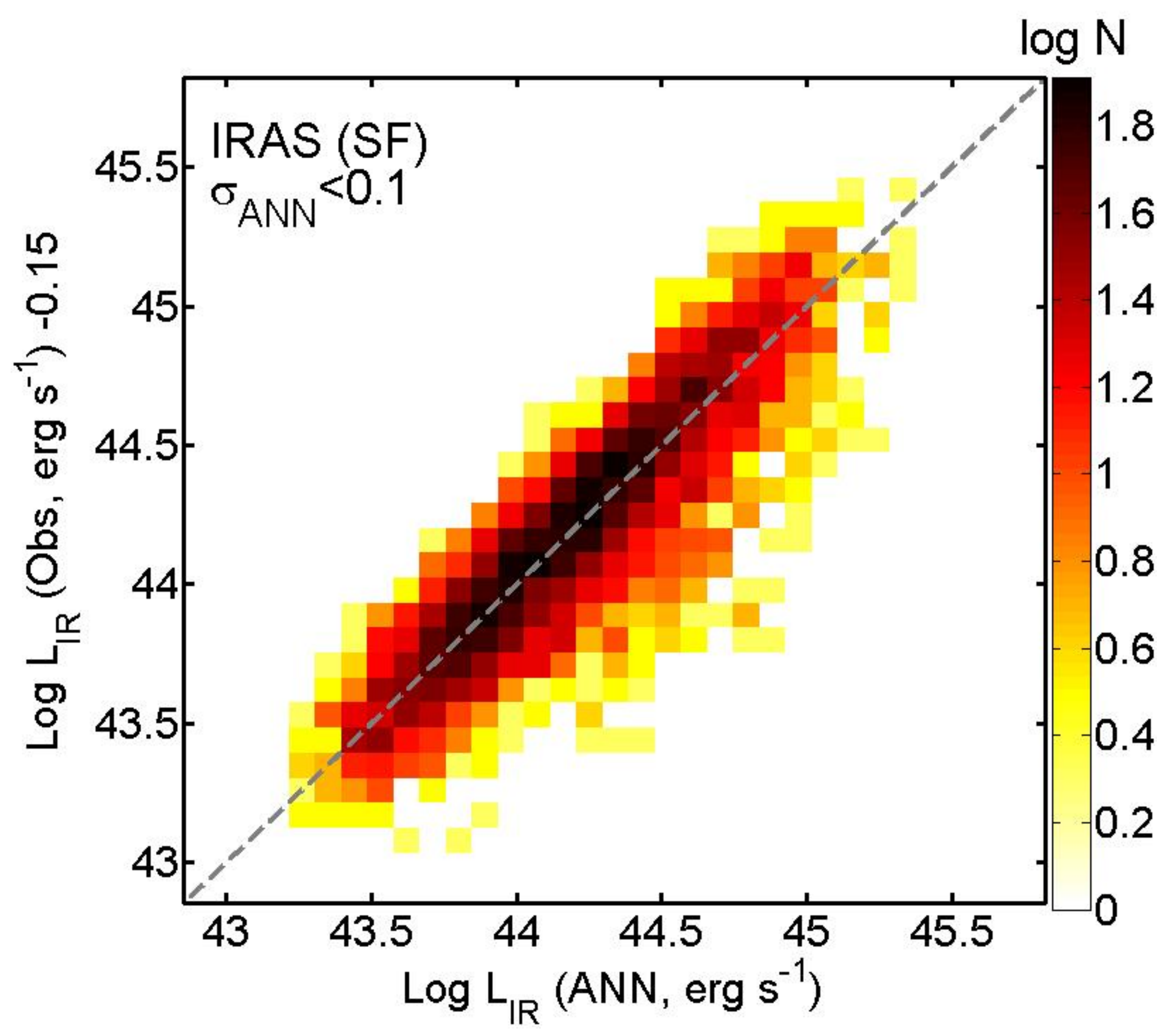}
\includegraphics[width=7.cm,height=5.5cm,angle=0]{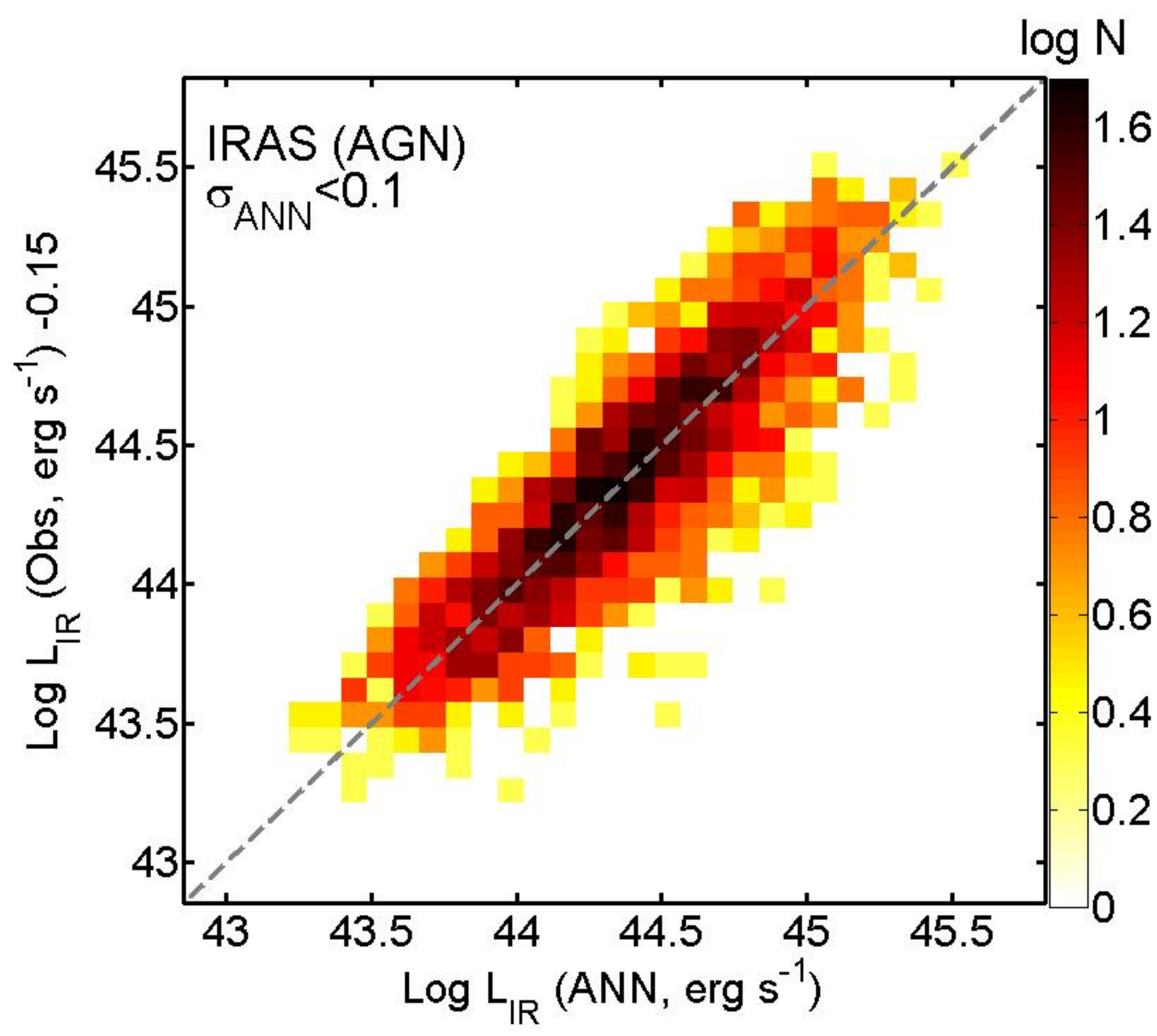}
\caption{A comparison between the observed and predicted \LIR\  for the IRAS sample
  after both a 0.15 dex offset has been applied to bring the values
  into line with the other catalogs and calibrations used in this paper,
  and after a 0.1 dex cut in ANN error has been applied.  The top,
  middle and lower panels show all, star forming and AGN galaxies respectively.
The scatter for the star-forming
and AGN galaxies is comparable: 0.20 and 0.24 dex, respectively.}
\label{fig-iras}
\end{figure}

In Figure \ref{fig-iras} we compare the ANN predictions with the
corrected IRAS \LIR\ for 7299 (out of a total of 8602 for which \LIR\ have been
predicted) galaxies that have $\sigma_{\rm ANN}<$ 0.1 dex.
The top panel shows the scatter for all galaxies, and the middle and lower
panels distinguish star-forming and AGN galaxies, analogous to the AKARI
validations shown in Figure \ref{fig-akari}.

Once the IRAS \LIR\ have been corrected for the known offset from AKARI,
Herschel and SDSS, the top panel of Figure \ref{fig-iras} demonstrates that
there is good agreement between the predicted and observed infra-red
luminosities, with a scatter of 0.24 dex.  The scatter amongst the star-forming
and AGN galaxies is 0.23 and 0.25 dex, respectively.  The performance
of the validation is therefore very similar for both the AKARI and IRAS
datasets.

\medskip

\begin{figure*}
\centering
\includegraphics[width=18cm,angle=0]{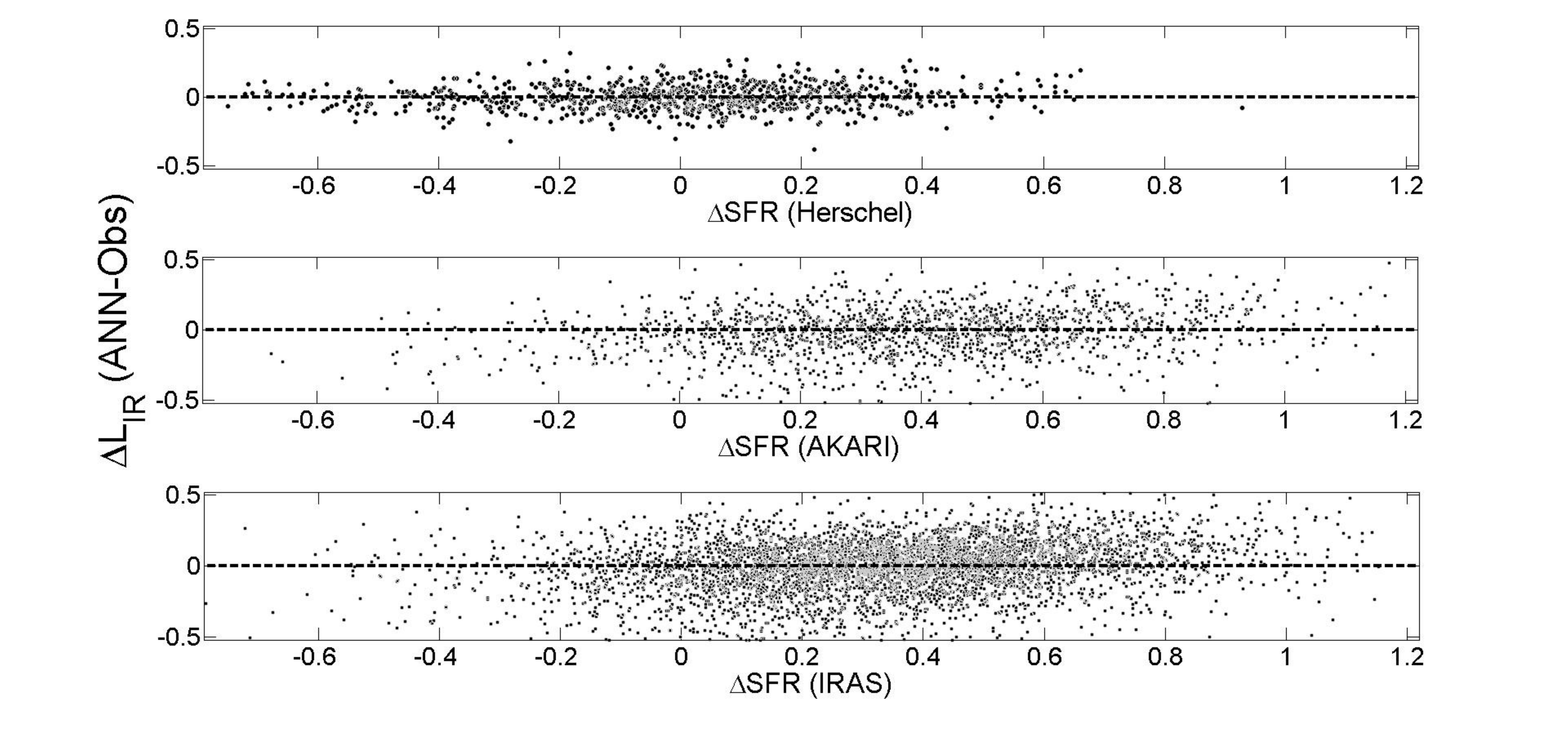}
\caption{The accuracy of the ANN \LIR\ prediction as a function of
$\Delta$ SFR, the offset of a galaxy from the star forming main
sequence.  In general, the ANN is able to robustly predict \LIR\
not only for galaxies on the main sequence ($\Delta$ SFR $\sim$ 0)
but also for outliers.  For (rare) galaxies with very enhanced SFRs
($\Delta$ SFR $>$ 0.8) there is tendency for the ANN to modestly
over-predict the \LIR, due to the lack of galaxies in the training
set in this regime. }
\label{delta_sfr}
\end{figure*}

As part of the validation process, we have also checked that $\Delta$ \LIR\
(the difference between the predicted and measured luminosities)
shows no residuals with any of the input parameters, and confirm that there is
no systematic trend or offset with any of the parameters listed in Table 
\ref{tab-input}.  As a final additional check, we confirm
that the scatter in $\Delta$ \LIR\ is not driven by the relative position of a galaxy
in the star forming main sequence (stellar mass versus star formation rate).
That is, we verify that we can equally well predict \LIR\ for galaxies that
have relatively high or low SFRs for their stellar mass, as well as for
those that are more typical.  In Fig. \ref{delta_sfr} for star-forming
galaxies only, we plot $\Delta$ \LIR\ as a function of $\Delta$ SFR (the
offset of a given galaxy from the main sequence, as described in Section
\ref{sec_compare}) for the Herschel, AKARI and IRAS samples.  It can be
seen that there is, in general, no correlation between $\Delta$ \LIR\
and $\Delta$ SFR, indicating that the scatter in the former is not
driven our inability to predict \LIR\ for main sequence outliers.
For the (rare) galaxies with very high SFR enhancements ($\Delta$ SFR $>$0.8,
or a factor of 6 above the main sequence) in AKARI and IRAS there is
a tendency of the ANN to modestly over-predict the \LIR; this is likely
due to the lack of such galaxies in the Herschel training set.
Remaining sources of scatter could be physical parameters not
included in our training set, uncertainties in observational measurements
are intrinsic scatter between \LIR\ and other parameters.

\section{Comparison with SDSS SFRs}\label{sec-sdss}

In the previous sections, we have described the training and
validiation of artificial neural networks that may be used to predict the
infra-red luminosities of galaxies.  We have demonstrated that the trained
networks perform well on independent datasets, that we are able to reject galaxies with
poorly predicted \LIR\ via the imposition of a $\sigma_{\rm ANN}$ threshold,
and that the \LIR\ predictions are well suited
for galaxies dominated by either star formation or AGN.

\begin{figure}
\centering
\includegraphics[width=9cm,angle=0]{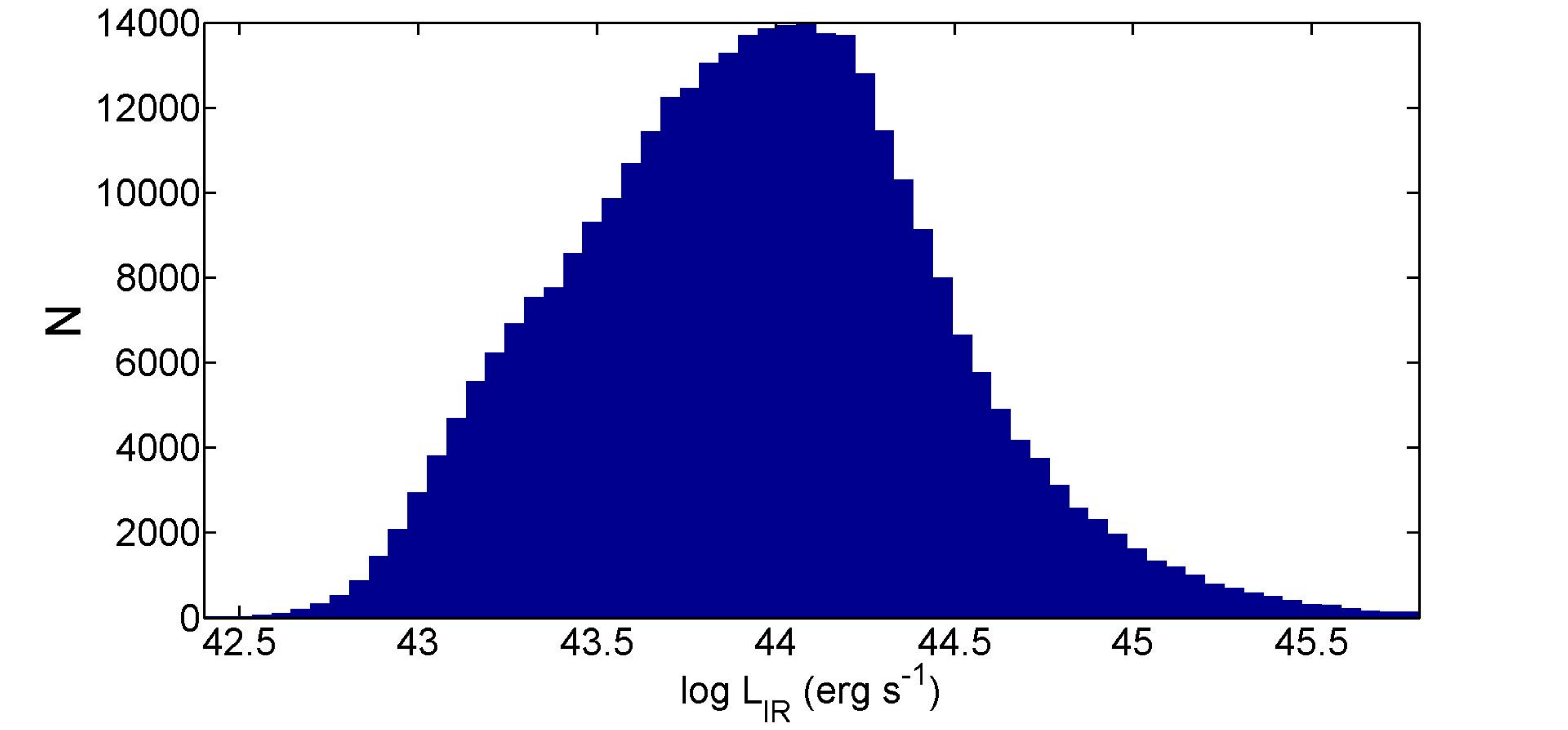}
\includegraphics[width=9cm,angle=0]{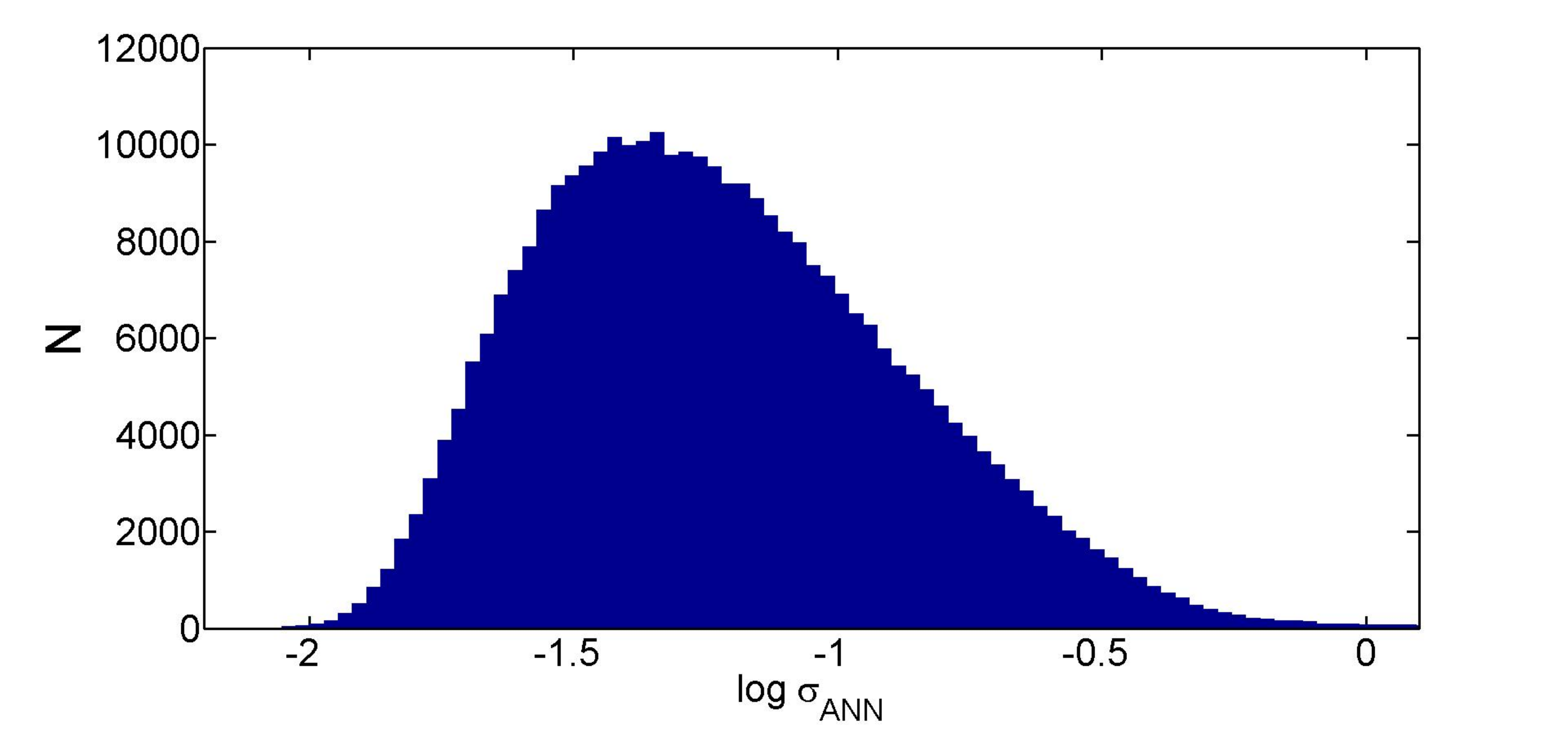}
\caption{The distribution of the ANN-derived \LIR\ values for SDSS galaxies and the associated errors.}
\label{fig-mu-her}
\end{figure}

We are now ready to apply the trained and validated networks to the
SDSS in order to predict their infra-red
luminosities. Specifically, we apply the trained ANN to the $\sim$ 332,000 galaxies
for which all requisite training parameters are available.
In Figure \ref{fig-mu-her} we show the distribution of the \LIR\ values
predicted for these
331,926 SDSS galaxies and the associated errors (described in Section \ref{error_sec}).  Of these galaxies, 128,507 are AGN.
The infra-red luminosities and associated
network error are presented in an online catalog that accompanies this paper. An
excerpt from the catalog, representing its first 10 lines, is given in Table
\ref{tab-catalog}.  For
completeness, all target galaxies are listed, but 247,137 SDSS galaxies pass the
$\sigma_{\rm ANN}<$0.1 dex criterion that we have adopted in this paper, of which
78,039 are classified as AGN (Kauffmann et al. 2003, S/N$\ge$ 1).

\begin{table*}
\begin{center}
\caption{Catalog of ANN predicted \LIR\ for SDSS galaxies.  The first 10 entries are
shown here; the full catalog is available in the online material that accompanies
this paper.}
\begin{tabular}{llllll}
\hline
SDSS objID &  RA  & Dec.  &  $z$  &  log \LIR\  & $\sigma_{\rm ANN}$ \\
 & (deg.) & (deg.)  & & (erg s$^{-1}$) &  \\
\hline
 587722953304310243 &  237.26196   &  0.25983     & 0.03464 & 43.34             & 0.07  \\  
 587722953304179097 &  236.88773   &  0.39001     & 0.11340 & 44.22             & 0.05    \\
 587722953304179083 &  236.88157   &  0.34449     & 0.03292 & 42.86             & 0.12    \\
 587722953304113519 &  236.77592   &  0.34773     & 0.03334 & 43.15             & 0.09    \\
 587722953303917041 &  236.36319   &  0.21511     & 0.10203 & 44.00             & 0.08    \\
 587722953303916807 &  236.22179   &  0.40430     & 0.03331 & 43.21             & 0.09    \\
 587722953304375633 &  237.28486   &  0.41628     & 0.19134 & 44.81             & 0.14    \\
 587722953304572300 &  237.81142   &  0.29300     & 0.03221 & 43.21             & 0.14    \\
 587722953304572365 &  237.86590   &  0.23560     & 0.08777 & 43.85             & 0.03    \\
 587722953304834385 &  238.37377   &  0.33674     & 0.09379 & 44.39             & 0.04    \\ \hline
\end{tabular}
\label{tab-catalog}
\end{center}
\end{table*}

\begin{figure*}
\centering
\includegraphics[width=18cm,angle=0]{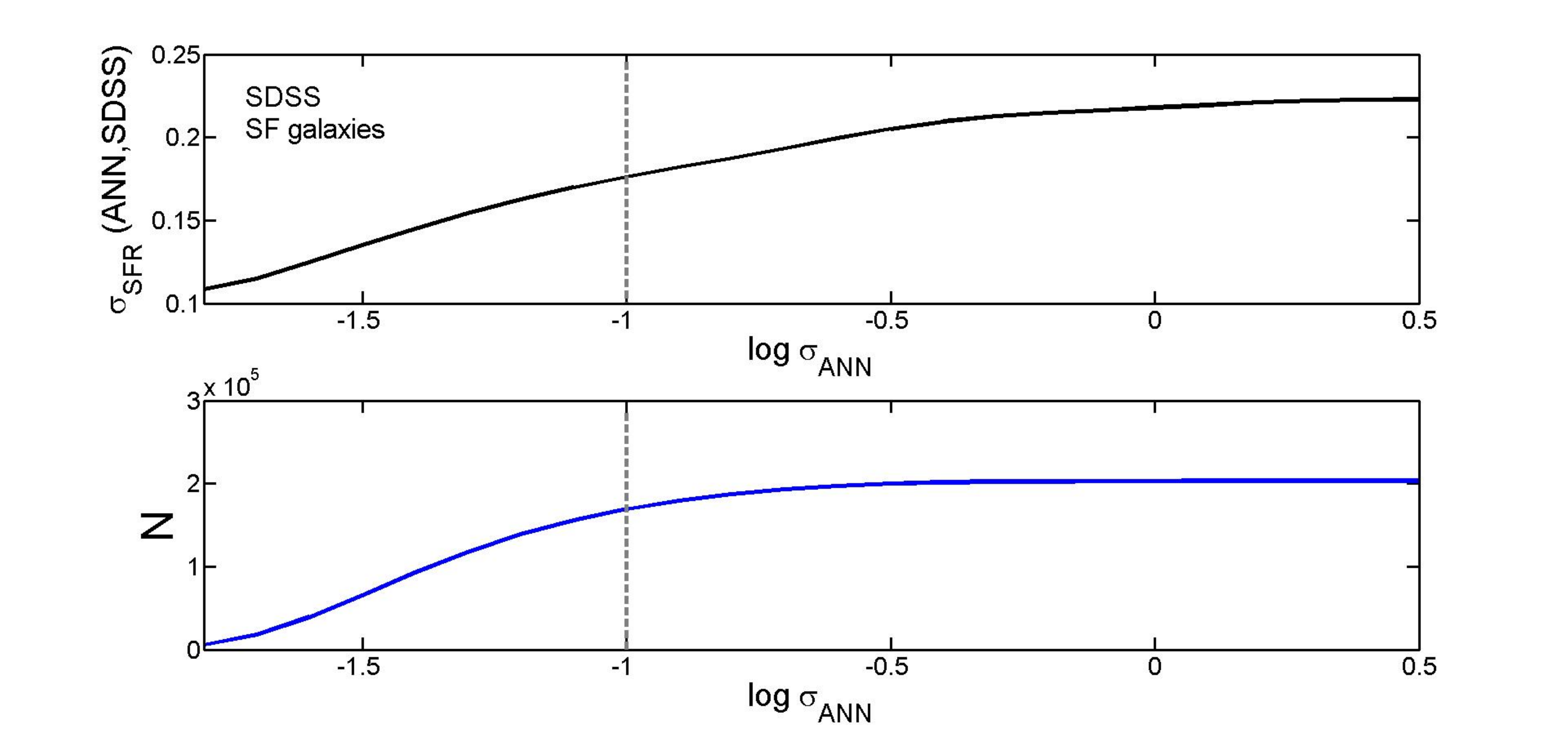}
\caption{The effect on sample size (lower panel) and scatter (upper panel) as a function
of $\sigma_{\rm ANN}$ threshold on the SDSS sample.  
The nominal requirement of $\sigma_{\rm ANN} < 0.1$ adopted for
this work is shown by a vertical dashed line in both panels.}
\label{sig_cut_sdss}
\end{figure*}

In order to compare the predicted \LIR\ from the ANN to the SFRs derived
from the SDSS spectra, we use Eqn. \ref{eqn-sfr} to convert the predicted
infra-red luminosities to star formation rates.  
In the top panel of Fig. \ref{sig_cut_sdss} we show 
the scatter in the difference between the SFR derived from the ANN-predicted
\LIR, and the SFR derived from aperture corrected SDSS spectra for star-forming
galaxies only.  As shown by Rosario et al. (2015), we expect these two SFR
indicators to be in good agreement.  In the lower panel of Fig. \ref{sig_cut_sdss}
we show how the sample size is affected by varying the $\sigma_{\rm ANN}$ threshold.  
Taken together, the two panels of Fig. \ref{sig_cut_sdss}
demonstrate that a cut at $\sigma_{\rm ANN}$=0.1 is a good balance between sample
size and accuracy.  Stricter cuts yield a smaller scatter in the difference
between SFR indicators, at the expense of sample size.

\begin{figure}
\centering
\includegraphics[width=7.cm,angle=0]{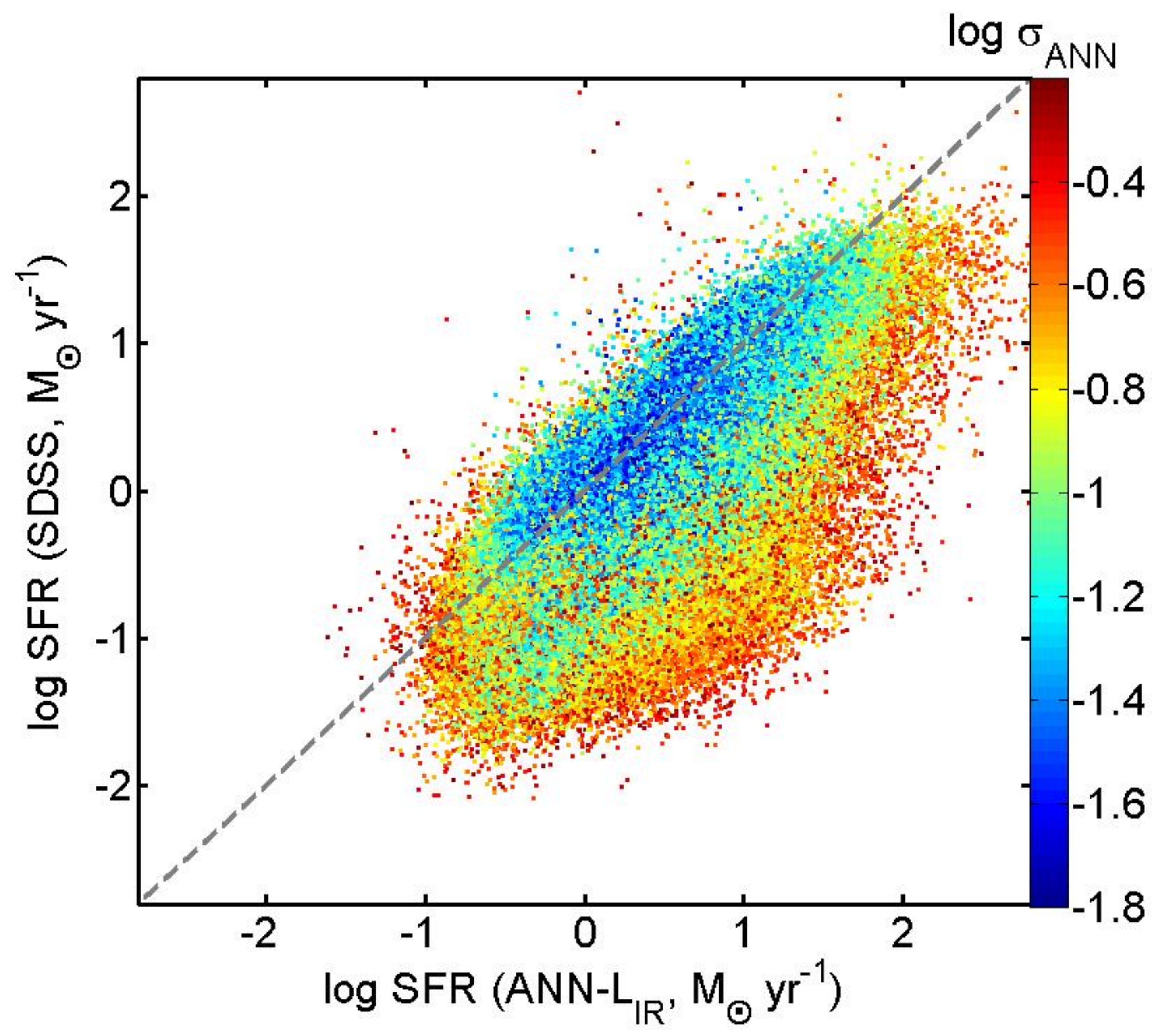}
\includegraphics[width=7.cm,angle=0]{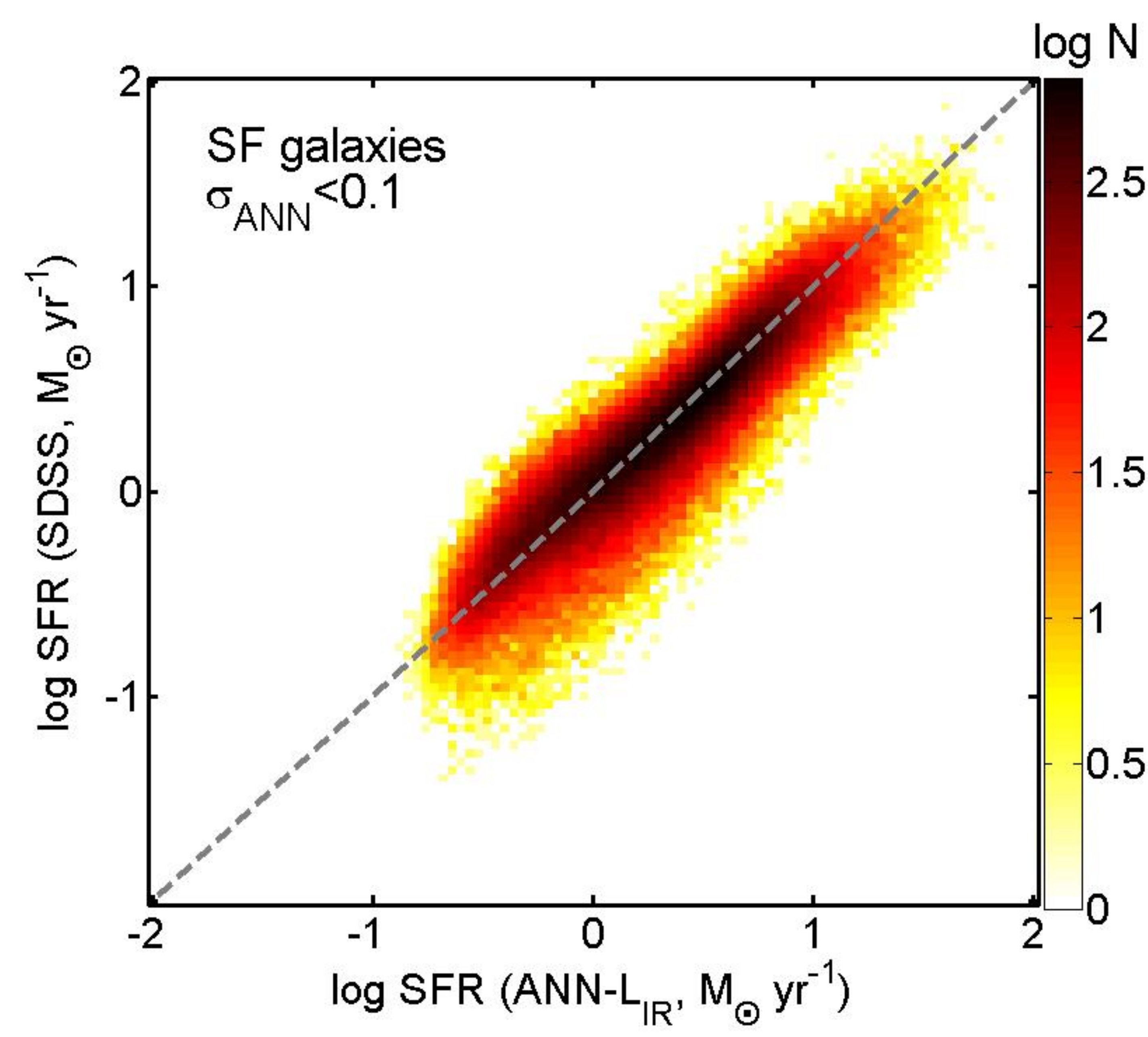}
\includegraphics[width=7.cm,angle=0]{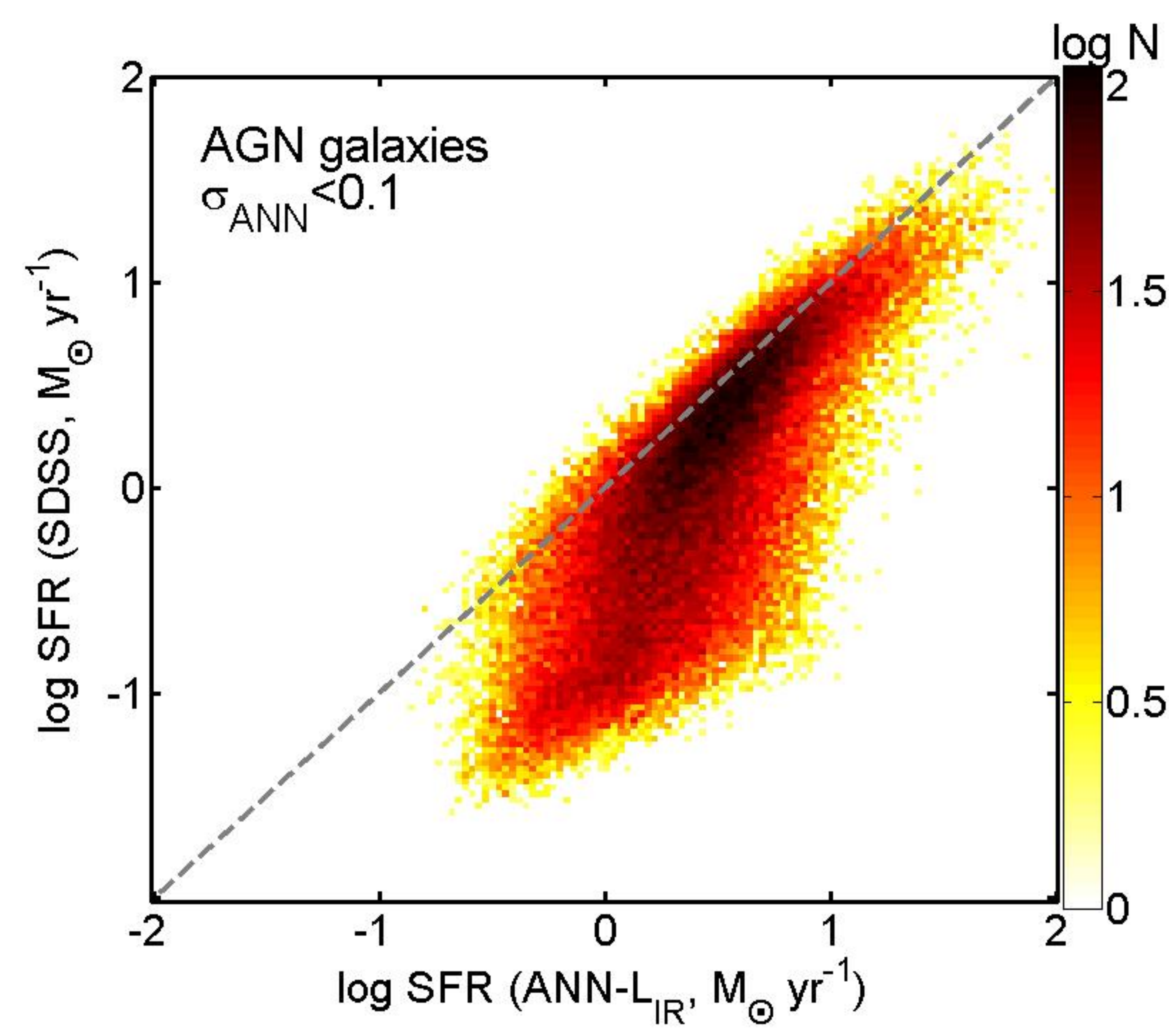}
\caption{SDSS SFRs compared to the SFRs (Eqn \ref{eqn-sfr}) derived for our predicted
  infra-red luminosities.  Top panel: SFRs for $\sim$ 332,000 galaxies with \LIR\
predictions,
with points colour colour coded by the ANN uncertainty ($\sigma_{\rm ANN}$), derived from the 
scatter amongst 20 trained networks.  In the middle and lower panels we show only
galaxies with $\sigma_{\rm ANN}<0.1$.  The colour coding represents  the number of galaxies 
in each cell.  The middle and lower panels show the SFRs for star-forming and AGN dominated
galaxies, respective.}
\label{fig-SFR-N}
\end{figure}

To further understand the impact of the $\sigma_{\rm ANN}$ criterion,
in the top panel of Figure \ref{fig-SFR-N} we compare the SFRs derived from our \LIR\ predictions to
all $\sim$ 332,000 SDSS galaxies with MPA/JHU star formation rates, with no
error requirement imposed.  The points are
colour-coded by the ANN error (recall, this is a metric of the confidence of the
ANN prediction based on the scatter from 20 trained networks).  The top panel
of the figure includes
all galaxies, regardless of emission line classification.  The majority of galaxies
follow the 1:1 line (dashed line), and these are also seen to be the cases with
the smallest associated ANN uncertainties.  Galaxies with larger $\sigma_{\rm ANN}$ yield
a larger scatter around the 1:1 line.  There is also a clear population of
galaxies with large $\sigma_{\rm ANN}$ that lie preferentially below the 1:1 line,
i.e. where the SDSS SFR is lower then the \LIR\ derived value by an order
of magnitude or more.

In the middle and lower panels of
Figure \ref{fig-SFR-N} we now impose a 0.1 dex error cut, and distinguish the star-forming
and AGN dominated galaxies (middle and lower panels respectively). The SFRs derived from
the \LIR\ for star-forming galaxies are seen to be in excellent agreement with those
derived from SDSS; the scatter is 0.17 dex for the 169,098 star forming galaxies 
that pass the error threshold (see also
Fig. \ref{sig_cut_sdss}).  However, even with the error cut,
the lower panel of Figure \ref{fig-SFR-N} shows a persistent disagreement between
the SDSS and ANN \LIR-based SFRs for the 78,039 AGN that pass the same
$\sigma_{\rm ANN}<$ 0.1 dex threshold, in the sense that SDSS SFRs are lower than those derived
from the \LIR\ by typically $\sim$ 0.5 dex (a factor of three).  The disagreement
between the methods increases as the SFRs decline, although the discrepancy
persists even at the highest SFRs of the sample.

The discrepancy shown in the lower panel of Figure \ref{fig-SFR-N} could be
indicative of a problem with the ANN predictions of \LIR\ in AGN galaxies.
However, the results presented in Section
\ref{sec-valid} demonstrate successful validation of the network with 3 different
IR datasets (Herschel, AKARI and IRAS).
The robustness of the network predictions in all our tests indicates that
it may be the SDSS SFRs that are under-estimated in many of the AGN galaxies
plotted in  Figure \ref{fig-SFR-N} (which, it should be noted, is only a subset of
all the AGN in the SDSS).
Indeed, an apparent under-estimate of SDSS SFRs compared to \LIR\ SFRs has
been recently reported by Rosario et al. (2015) for the AGN, composite
and low SFR galaxies in the original Herschel Stripe 82 data (see also
Fan et al. 2013 for a similar result in the H-ATLAS SDP data).
Rosario et al. (2015) show that there is an increasing disagreement between
Herschel and SDSS SFRs as the error on the latter increases, indicating
that the discrepancy is an Eddington bias caused by the limit in Herschel's
sensitivity.  With a deeper IR dataset, we would expect to see a symmetric scatter in
the comparison between SDSS galaxies with large errors and Herschel, but
with the current IR detection limits, we only see one side of this distribution
(galaxies on the other side of the distribution are not detected by Herschel).
This scenario is confirmed by Rosario et al. (2015) with a Monte Carlo simulation of
the SDSS error distribution and a comparison of the Herschel detection limit.

 \begin{figure}
\centering
\includegraphics[width=7.cm,height=5.5cm,angle=0]{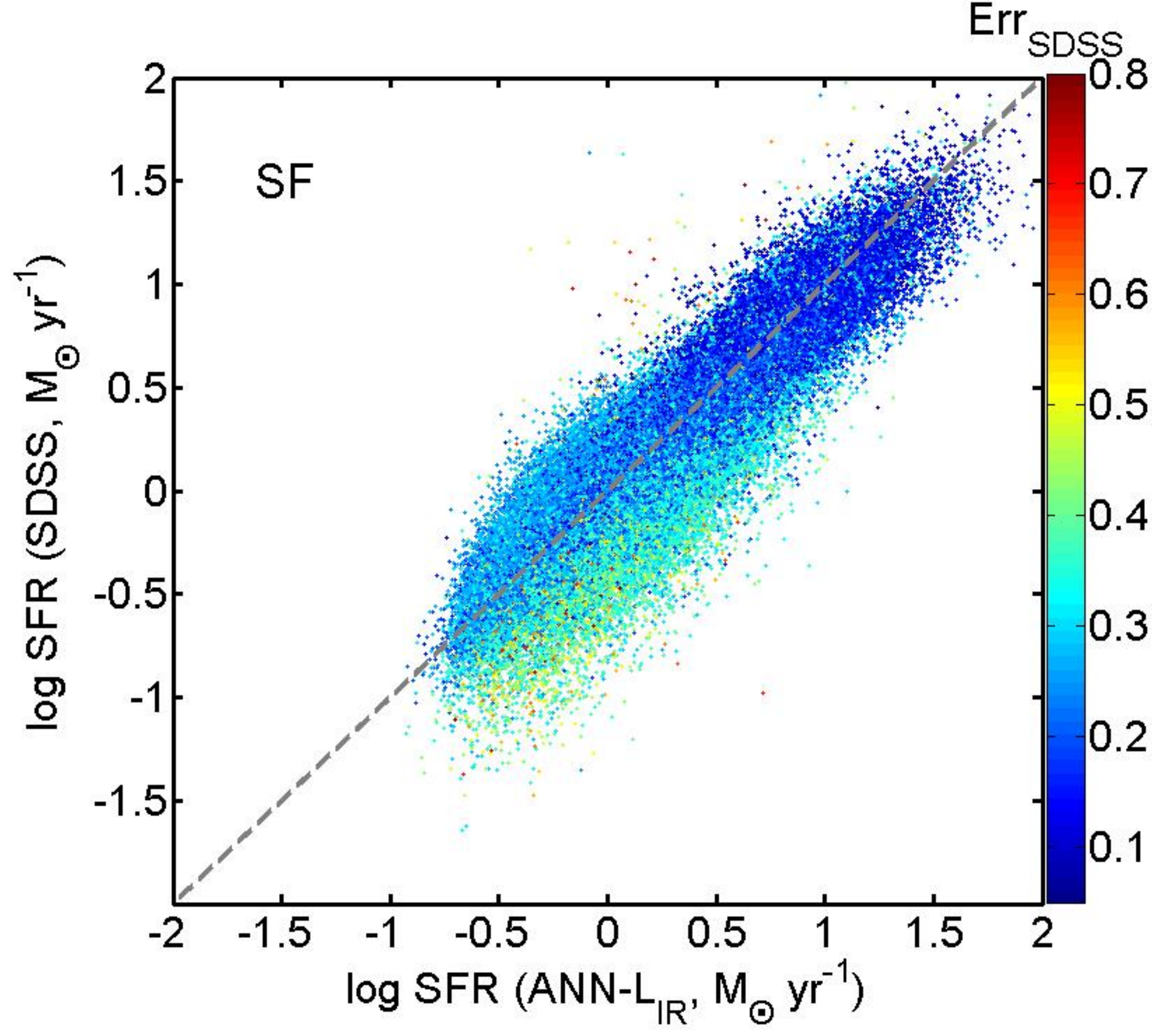}
\includegraphics[width=7.cm,height=5.5cm,angle=0]{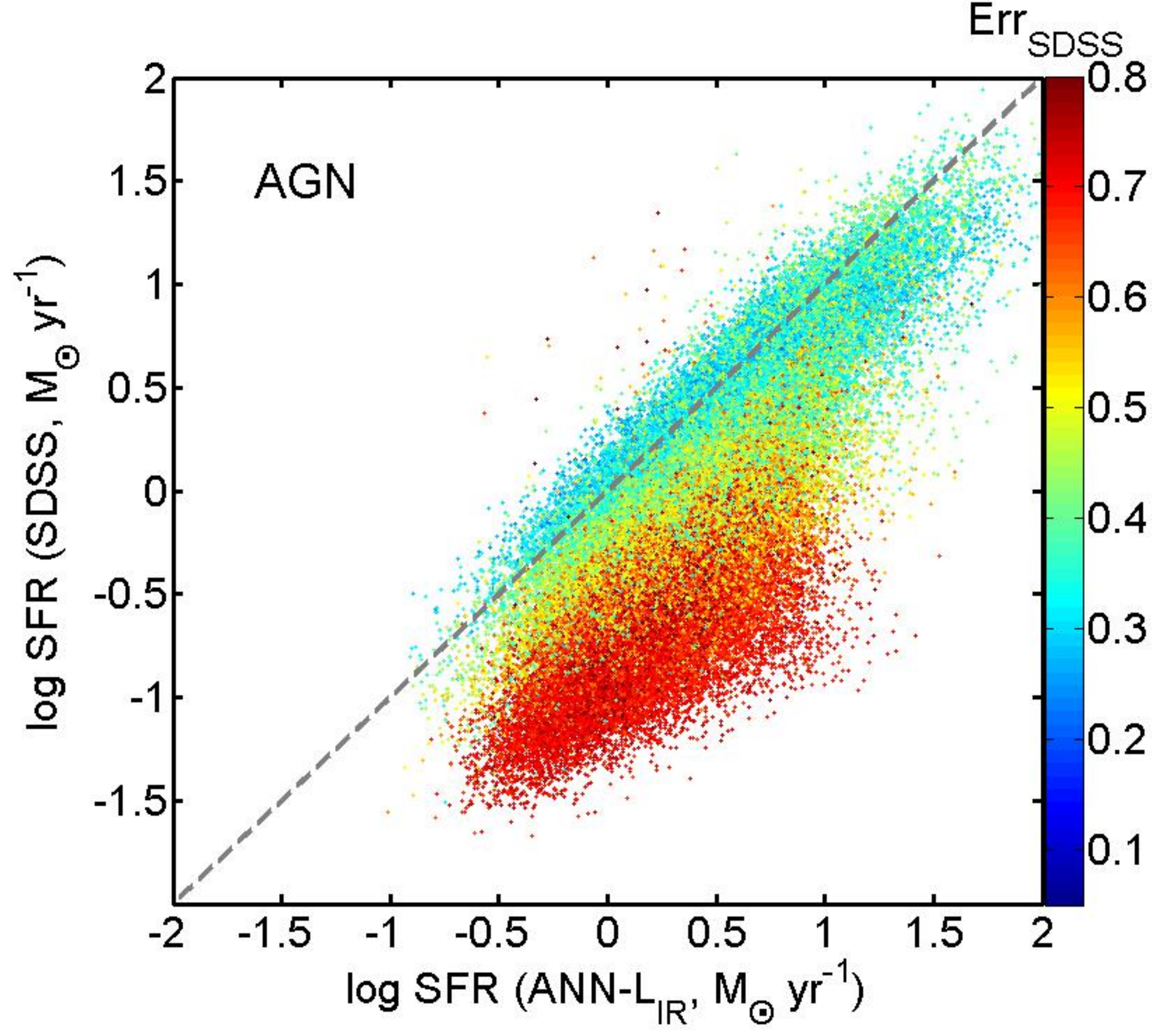}
\caption{The same as Figure \ref{fig-SFR-N} for star forming and AGN galaxies (upper
  and lower panels respectively), but now colour coded by the error in the SDSS
SFR.}
\label{fig-err-cut}
\end{figure}

The results of Rosario et al. (2015) suggest that
the offset between SDSS SFRs and those derived through ANN predictions
of the \LIR\ for AGN galaxies (lower panel of Figure \ref{fig-SFR-N}) is likely
also to be due to the uncertainties in the former and the detection threshold of
the latter.  We test this hypothesis 
in Figure \ref{fig-err-cut}, where we show again the comparison between SFRs determined
from our \LIR\ ANN predictions and those from the SDSS.  These are the same
galaxies shown in Fig. \ref{fig-SFR-N}, but now colour coded by the error on
the SDSS SFR.  The star forming galaxies, which we have already shown in the
middle panel of Fig. \ref{fig-SFR-N} exhibit a universally good agreement between
the ANN and SDSS, all have relatively small SFR errors, typically $<$0.3 dex.
However, there is a large spread in the SDSS SFR errors in the AGN population
(Fig. \ref{fig-err-cut}, lower panel).  Approximately 7 per cent 
of AGN have errors $<$ 0.3 dex, the median error is 0.5 dex
and a significant number of galaxies with AGN have errors as large as 0.7--0.8 dex.  
A striking result
of Fig. \ref{fig-err-cut} is that as the error in the SDSS SFR increases, the
disagreement with the ANN SFR also increases, and the offset is very similar
to the error in the SDSS SFR.  This demonstrates that the errors on the
SDSS SFRs have been well determined.  The distribution of points in the lower panel of
Fig. \ref{fig-err-cut} is therefore the result of the same Eddington bias
seen in the Herschel training set, that leads to an apparent one-sided scatter in the
comparison as the error in the SDSS SFR increases.  Galaxies that scatter
above the 1:1 line are not detected by the Herschel training set, and therefore
do not have ANN predictions made.  We therefore conclude that the SFRs
from the ANN are robust (as supported by our validation tests),
and that the discrepancy with the SDSS values is due to the typically
large uncertainties in the latter.
Our results and those of Rosario et al. (2015) highlight the niche of FIR-based SFRs,
which can yield accurate values in regimes where the optical determinations
have large associated errors.

\section{Summary}

Based on a sample of 1136 galaxies identified in a cross-match between
the SDSS and Herschel Stripe 82 Survey, we have trained an artificial
neural network to predict IR luminosities based on 23 input parameters
measured from SDSS imaging and spectroscopy.  The scatter in the predicted
\LIR\ for a self-validation test is $<$0.1 dex for both star-forming and 
AGN dominated galaxies.
The network performance is validated on two independent datasets: the IRAS
and AKARI all sky surveys, which have both been matched to SDSS galaxies.
The scatter between the predicted and observed \LIR\ for both surveys is
$\sim$0.25 dex, after imposing an uncertainty threshold of $\sigma_{\rm ANN}<0.1$.
We demonstrate that the ANN is able to  predict \LIR\ for both star-forming
and AGN dominated galaxies with similar scatter.  The ANN is also able
to accurately predict the \LIR\ (and hence SFR) even for galaxies lying significantly off
the star-forming main sequence.

The trained networks are applied to $\sim$ 332,000 SDSS galaxies for which
all 23 input parameters are available, including $\sim$ 129,000 AGN.  Of these
$\sim$ 332,000 galaxies, 247,137 have $\sigma_{\rm ANN} < 0.1$ dex, including $\sim$ 78,000 AGN.  
We provide a catalog of the ANN predicted infra-red
luminosities and their uncertainties ($\sigma_{\rm ANN}$).
The SFRs inferred from the ANN \LIR\
predictions are in excellent agreement with SDSS spectroscopic SFRs for
star-forming galaxies.  However, for the AGN with \LIR\ predictions, the SDSS SFR are apparently
under-estimated by up to an order of magnitude, even when the uncertainty threshold has been
imposed.  It is demonstrated that this apparent under-estimate is due to larger uncertainties
in the SDSS SFRs for AGN galaxies, which are based on D$_{4000}$, and the detection
threshold in the training set. 
The \LIR\ predictions provided here can therefore be used to obtain more accurate
SFR measurements in AGN, or in weak emission galaxies where there exist
large uncertainties in the SDSS derived values (Fan et al. 2013; Rosario et al. 2015).

\section*{Acknowledgments} 

We are grateful to Ho Seong Hwang for providing his catalogs of AKARI
and IRAS fluxes for matched SDSS galaxies.  SLE acknowledges the receipt
of an NSERC Discovery Grant.

Funding for the SDSS and SDSS-II has been provided by the Alfred
P. Sloan Foundation, the Participating Institutions, the National
Science Foundation, the U.S. Department of Energy, the National
Aeronautics and Space Administration, the Japanese Monbukagakusho, the
Max Planck Society, and the Higher Education Funding Council for
England. The SDSS Web Site is http://www.sdss.org/.

The SDSS is managed by the Astrophysical Research Consortium for the
Participating Institutions. The Participating Institutions are the
American Museum of Natural History, Astrophysical Institute Potsdam,
University of Basel, University of Cambridge, Case Western Reserve
University, University of Chicago, Drexel University, Fermilab, the
Institute for Advanced Study, the Japan Participation Group, Johns
Hopkins University, the Joint Institute for Nuclear Astrophysics, the
Kavli Institute for Particle Astrophysics and Cosmology, the Korean
Scientist Group, the Chinese Academy of Sciences (LAMOST), Los Alamos
National Laboratory, the Max-Planck-Institute for Astronomy (MPIA),
the Max-Planck-Institute for Astrophysics (MPA), New Mexico State
University, Ohio State University, University of Pittsburgh,
University of Portsmouth, Princeton University, the United States
Naval Observatory, and the University of Washington.

\end{document}